# Linear Theory of Visco-Resistive Tearing Instability

Tohru Shimizu

*Research Center for Space and Cosmic Evolution, Ehime University, Matsuyama 790-8577, Japan*

March 2, 2024

In this study, a new linear theory of tearing instability is shown, where the modified LSC (Loureiro, Schekochihin, and Cowley) theory [36] developed from the original LSC theory [8] is extended from inviscid-resistive MHD to viscous-resistive MHD. In contrast to FKR [2] and original LSC theories, the upstream open boundary condition is implemented at a finite point $\xi_c$, which is an additional control parameter to determine the solutions. This paper firstly studies when the resistivity and viscosity are uniform in space. In addition, some variations in the nonuniformity are studied. It is shown that the non-uniformity can enhance the linear growth rate, rather than uniform case. Unexpectedly, this suggests that the forward cascade process of plasmoid instability (PI) does not stop, i.e., the finite differential MHD simulations fail. To stop the forward cascade, uniform viscosity is required not only in the inner region of the current sheet but also in the outer region. In the uniform case, the critical condition is predicted to be $2P_m/(S\xi_c) = 0.06$, beyond which the tearing instability, i.e., the forward cascade, stops. Here, $S$ is the Lundquist number, $P_m$ is the magnetic Prandtl number, and $\xi_c$ is the distance between the upstream open boundary and neutral sheet, where the current sheet thickness is fixed at $\xi_0 = 1.307$. According to the critical condition, the resistivity and viscosity employed in most MHD simulations of PI are too small to stop the forward cascade. This critical condition may be also applicable for the trigger problem of the current sheet destabilization in substorms and solar flares.

## 1   Introduction

The magnetic reconnection process is an energy conversion mechanism to convert magnetic energy to plasma kinetic energy. Regarding the explosive energy conversion observed in solar flares and geomagnetic substorms, the magnetic reconnection process must be fast. The fast magnetic reconnection process has been studied during the past 60 years [1,2,3,4]. In basic plasma physics, it is considered that the fast magnetic reconnection process starts from the collapse of the current sheet, which is called tearing instability [2]. Hence, to attain fast magnetic reconnection, tearing instability may be required to be quickly caused. Studying fast tearing instability is important to explore the fast magnetic reconnection process.

FKR theory (Furth, Killeen, and Rosenbluth [2]) was proposed in the 1960s to explore the linear growth stage of the tearing instability. Then, numerous linear studies of the tearing instability have been reported until today. In most of those linear theories, the tearing instability was studied on the basis of the ideal-MHD limit equilibrium, i.e., zero-resistivity limit [2,5,6,7]. In fact, the assumed

*author's e-mail: shimizu@cosmos.ehime-u.ac.jp*

equilibrium is satisfied only when the resistivity is zero. At this point, if the resistivity is zero, magnetic reconnection process does not occur, and hence, tearing instability does not occur. In other words, since a finite resistivity violates the assumed equilibrium, the linear theory based on the inaccurate equilibrium has delicate problems. At least, such linear theories will be inapplicable when the magnetic diffusion speed based on the resistivity is close to the Alfven wave speed. Notably, when the current sheet is thin, those speeds may be close, in which fast tearing instability may occur.

LSC theory (Loureiro, Schekochihin, and Cowley [8]) largely improved the equilibrium problem in FKR theory, where the nonzero-flow equilibrium field was employed, by which the equilibrium is exactly satisfied even when the resistivity is large. In this paper, we refer to this LSC theory as the original LSC theory. Similar to FKR theory, the original LSC theory "analytically" solves the linear perturbation equations between the neutral sheet ($\xi = 0$) and the infinity upstream point ($\xi = +\infty$). In addition, the region outside the current sheet is simply assumed to be an ideal MHD, and hence, a finite resistivity works only inside the current sheet. The assumption of the ideal MHD outside the current sheet is often employed in linear studies of the tearing instability [2,5,6,8,9,10,11,12,13]. This means that the resistivity is assumed to be non-uniform in space. It is believed that the case of non-uniform resistivity is not essentially different from the case of uniform resistivity. Certainly, since the magnetic reconnection process requires some form of resistivity inside the current sheet to reconnect magnetic field lines, the outside area may not need to have a finite resistivity. However, it will be worth rigorously studying whether the non-uniform case is exactly the same as the uniform case or not.

In fact, in the numerical MHD simulations of the fast magnetic reconnection process, the selection of a uniform resistivity or non-uniform resistivity has been a historically controversial topic. As shown in many numerical MHD studies, non-uniform resistivity, such as anomalous resistivity, results in a fast magnetic reconnection known as the Petschek (PK) model [3,14,15,16,17,18,19]. Meanwhile, when the resistivity is assumed to be uniform, generating the PK model is a delicate problem [16,18,19,23]. If the uniform resistivity can reproduce the PK model, the non-uniformity of the resistivity is not important to make the magnetic reconnection fast. Inversely, if the uniform resistivity cannot reproduce the PK model, how non-uniform resistivity can reproduce it must be studied, which may have to be extensively studied in kinematic plasma physics.

On the other hand, according to some MHD simulations [20,25,26,27,28,29,30,31,32,33], uniform resistivity results in a turbulent Sweet-Parker (SP) model [1], which may cause fast magnetic reconnection. In such a case, we may again say that the nonuniformity of resistivity is not important to make the magnetic reconnection fast. However, if the uniform resistivity cannot cause a fast magnetic reconnection, issues similar to those of the PK model may occur.

Hence, in both the PK and PI models, it is important to determine whether fast magnetic reconnections can be achieved with "uniform" resistivity or not. The fast magnetic reconnection should be considered a nonlinear process that may have to be studied only in MHD simulations. At this point, since only the linear process is considered in this paper, such a nonlinear fast magnetic reconnection process cannot be studied. However, the uniformity of resistivity in the linear process of tearing instability is worthy of primary study.

To examine when the uniform resistivity is assumed not only inside of the current sheet but also outside, Shimizu modified the original LSC theory. Then, he named it the modified LSC theory [34,35,36, 37,38,39, 40,41]. The linear perturbation equations solved in the modified LSC theory are basically the same as that of the original LSC theory [8], but the resistive MHD region was extended from the inner region to the outer region. In addition, a kind of upstream open boundary condition was assumed at a finite upstream point. Then, how the tearing instability is affected by the upstream open boundary was studied. In addition, between the modified and original LSC theories, there is a difference in the normalizations of space and time. In fact, the linear growth rate $\lambda$ examined in the modified LSC theory is normalized by the wavelength $l_{cs}$ of the plasmoid chain. Meanwhile, the linear growth rate $\gamma \tau_A$ examined in the original LSC is normalized by the macroscopic Sweet-Parker (SP) sheet length $L_{cs}$. As explained in Fig. 22 of Shimizu's paper [36], $\gamma \tau_A$ can largely exceed unity, but $\lambda$ does not exceed unity. This means that tearing instability essentially occurs in the sub-Alfvenic time scale.

In contrast to FKR and the original LSC theories, the eigenvalue problem (EVP) solver, which is widely employed in the boundary value problem, is not used in the modified LSC theory. Instead, the modified LSC theory numerically solves initial value problems (IVPs). Since IVP is much simpler than EVP, the solutions numerically obtained in IVP will have much higher numerical precision. For example, the linear perturbation equations of IVP are discretized in $10000 \sim 100000$ elements, i.e., numerical grid size $\delta \xi = 0.0001 \sim 0.00001$ between the neutral sheet ($\xi = 0$) and upstream boundary ($\xi = \xi_c$). In contrast, the number of elements in the EVP solver is generally much lower. Hence, using IVP will be more advantageous than using EVP solvers.

The main theme of this paper is to introduce the viscosity in the modified LSC theory of tearing instability in a simple slab geometry, where the critical condition is derived, below which the current sheet is stabilized, i.e., tearing instability does not occur. Then, it is shown that the critical condition depends on the location of the upstream open boundary, in addition to the resistivity, and viscosity. There are many previous studies in which the viscosity was introduced in the linear theory of tearing instability [5,6,9,11,12,13,21,22,24]. However, there is no study in which the upstream boundary condition is examined and the critical condition is obtained. At the end of this paper, it is shown that viscosity, which is uniformly assumed not only inside but also outside the current sheet, effectively works to stabilize the current sheet, and hence can stop the tearing instability.

In Chapter 2, the linear perturbation equations to be solved in this paper are introduced. These equations are derived from incompressible MHD equations. Five cases, in which the spatial non-uniformities of the resistivity and viscosity are considered, are studied. In Chapter 3, the numerical IVP techniques used to find the solutions are explained. In Chapter 4, the numerical results are shown, and the critical conditions of $\lambda = 0$ are derived in each case. The results of Case 2 are similar to those obtained with the FKR and original LSC theories. Rather, Case 3 is the highlight of this paper. In Chapter 5, a discussion and applications are presented. Finally, in Chapter 6, a summary is provided.

## 2   Linearized MHD equations

### 2.1   2D equilibrium flow field in a 1D current sheet

In this chapter, some variations of the linear perturbation equations solved in this paper are derived from the viscous-resistive and incompressible MHD equations shown below.

$$\partial_t \mathbf{u} + \mathbf{u} \cdot \nabla \mathbf{u} = -\nabla \mathbf{P} + (\nabla \times \mathbf{B}) \times \mathbf{B} + \nu \nabla^2 \mathbf{u} \tag{1}$$

$$\partial_t \mathbf{B} = -\nabla \times (\mathbf{u} \times \mathbf{B}) + \eta \nabla^2 \mathbf{B} \tag{2}$$

In the 2D case of $\partial_z = 0$, $\mathbf{u}$ and $\mathbf{B}$ can be translated to $\phi$ and $\psi$ by $(u_x, u_y, u_z) = (-\partial_y \phi, \partial_x \phi, 0)$ and $(B_x, B_y, B_z) = (-\partial_y \psi, \partial_x \psi, 0)$. Here, $\nabla^2 = \partial_x^2 + \partial_y^2$. $\eta$ and $\nu$ represent the resistivity and viscosity, respectively, and they are basically uniform in space and constant in time. However, those values in Cases 2, 4, and 5 are set to be non-uniform. Eqs. (1) and (2) can be translated as follows.

$$\partial_t \nabla^2 \phi + (\partial_x \phi)(\partial_y \nabla^2 \phi) - (\partial_y \phi)(\partial_x \nabla^2 \phi) = (\partial_x \psi)(\partial_y \nabla^2 \psi) - (\partial_y \psi)(\partial_x \nabla^2 \psi) + \nu(\nabla^2)^2 \phi \qquad (3)$$

$$\partial_t \psi + (\partial_x \phi)(\partial_y \psi) - (\partial_y \phi)(\partial_x \psi) = \eta \nabla^2 \psi + E_0 \qquad (4)$$

First, we consider the equilibrium magnetic field of a 1D current sheet for $(B_x, B_y, B_z) = (0, B_{0y}(x), 0)$. We assume that the outer edge of the current sheet is located at $x = \delta_{cs}\xi_0$ and that the Alfven speed at the point is $V_A$, where $\delta_{cs}$ is the thickness of the current sheet. Hence, the outer edge of the current sheet is not at $x = \delta_{cs}$, where $\xi_0 = 1.307$ is set through this paper. In addition, we assume the 2D equilibrium flow field to be $\phi_0 = \Gamma_0 xy$, where $\Gamma_0 = 2V_A/L_{cs}$ and $L_{cs}$ is the current sheet length for the Sweet-Parker steady-state model. In the model, $\delta_{cs}$ and $L_{cs}$ are related with the Lundquist number $S = L_{cs}^2/(2\delta_{cs}^2) = V_A L_{cs}/\eta$. Furthermore, the equilibrium magnetic field satisfies the next equation.

$$\eta \partial_x^2 \psi_0 + \Gamma_0 x \partial_x \psi_0 + E_0 = 0 \qquad (5)$$

This equation is translated by $x = \delta_{cs}\xi$ as follows:

$$\eta \psi_0'' + \xi \psi_0' + E_0/\Gamma_0 = 0 \qquad (6)$$

Hereafter, the prime is the derivative of $\xi$. This equation can be solved as follows:

$$\psi_0' = f(\xi) = (-E_0/\Gamma_0)e^{-\xi^2/2} \int_0^\xi dz e^{z^2/2} \qquad (7)$$

Second, to normalize the current sheet configuration, we set $\psi_0' = \delta_{cs}B_{0y}(x) = \delta_{cs}V_A = 1$ and $\psi_0'' = 0$ at $\xi = \xi_0$, where the current density is zero. Then, we obtain $E_0 = -\Gamma_0\xi_0$. Hereafter, for the simplicity of expression, $\psi_0'$ is replaced by $f(\xi)$. To study the perturbation theory of tearing instability, we employ $\phi_0$ and $\psi_0$ as the rigorous equilibrium. Then, we set $\phi = \phi_0 + \delta\phi$, $\psi = \psi_0 + \delta\psi$, where $\delta\phi(x, y, t) = \phi_1(x, t)e^{ik(t)y}$ and $\delta\psi(x, y, t) = \psi_1(x, t)e^{ik(t)y}$ with $k(t) = k_0 e^{-\Gamma_0 t}$. Moreover, we set $\phi_1 = -i\Phi(x)e^{\gamma t}$ and $\psi_1 = \Psi(x)e^{\gamma t}$. Then, we translate Eqs. (3) and (4) with $x = \delta_{cs}\xi$, $\kappa = k_0 V_A/\Gamma_0 = k_0 L_{cs}/2$, $\epsilon = 2\delta_{cs}/L_{cs}$, and the linear growth rate $\lambda = \gamma/(\Gamma_0\kappa)$. In addition, we set $N = 2\nu/(V_A L_{cs}) = \nu/\Gamma_0$. As a result, the resistivity $\eta$ and viscosity $\nu$ are respectively translated to $\epsilon$ and $N$. On the basis of the setup shown above, the perturbation equations to be solved in this paper are derived. In the next section, five variations of the perturbation equations, as shown in Table 1, are studied. The details of the derivations are shown in Appendix.

## 2.2 Case 1: Uniform Resistivity and No Viscosity

In this case, the resistivity $\epsilon(> 0)$ is uniform in space, and the viscosity $N$ is also uniformly zero, i.e., inviscid. This case is similar to the previous study on the modified LSC [36] theory, but $f(\xi) = 1$ is not employed in $\xi > 1.307$ in this paper. Instead, Eq. (7) is employed in $0 < \xi < \xi_c$, where $\xi_c$ is the location of the upstream boundary point. Hence, the equilibrium is rigorously established in all ranges of $0 < \xi < \xi_c$. The linear perturbation equations to be solved are as follows.

$$\lambda\Phi'' - \lambda\kappa^2\epsilon^2\Phi = -f(\xi)(\Psi'' - \kappa^2\epsilon^2\Psi) + f''(\xi)\Psi \qquad (8)$$

$$\Psi'' - \kappa^2\epsilon^2\Psi = \kappa\lambda\Psi - \kappa f(\xi)\Phi. \qquad (9)$$

These equations have been derived by Loureiro et al. [8]. Solving these equations as IVP in $0 < \xi < \xi_c$, zero-crossing solutions that satisfy $\Phi = \Psi = 0$ at $\xi_c$ are obtained. Note that $\xi_c$ changes, which is indirectly a control parameter that can be used to explore the behaviors of the solutions. Fig. 1 shows the typical profile of the solutions obtained for $\kappa = 1$, $\epsilon = 0.1$, $N = 0$, $\lambda = 0.13$, and $\Phi'(0) = 3.0802$. This is a zero-crossing solution that satisfies $\Phi = \Psi = 0$ at $\xi_c = 2.06$. Finding this solution is explained in the next chapter.

| Case | Inner region ($\xi \leq 1.307$) | Condition at $\xi = 1.307$ | Outer region ($1.307 < \xi$) | limit | Type of Solution |
|---|---|---|---|---|---|
| Case1 | (8) (9) | — | same as inner region | — | Zero-Crossing |
| Case1a | (10) (11) | — | same as inner region | $\lambda = 0$ | Zero-Crossing |
| Case1b | (12) (13) | — | same as inner region | — | Zero-Crossing |
| Case2 | (8) (9) | — | (12) (13) | — | Zero-Crossing |
| Case2a | (10) (11) | — | (11) (14) | $\lambda = 0$ | Zero-Crossing |
| Case3 | (15) (16) | — | same as inner region | — | Zero-Contact |
| Case3a | (17) (18) | — | same as inner region | $\lambda = 0$ | Zero-Contact |
| Case4 | (15) (16) | (19) | (8) (9) | — | Zero-Crossing |
| Case4a | (17) (18) | (11) (20) | (10) (11) | $\lambda = 0$ | Zero-Crossing |
| Case4b | (16) (19) | (8) | (8) (9) | $N = +\infty$ | Zero-Crossing |
| Case4c | (18) (19) | (11) (20) | (10) (11) | $\lambda = 0$ & $N = +\infty$ | Zero-Crossing |
| Case5 | (15) (16) | — | (8) (9) | — | Zero-Contact |
| Case5a | (17) (18) | $\Phi(\xi_0 - 0) = \Phi(\xi_0 + 0)$ | (11) (18) | $\lambda = 0$ | Zero-Contact |

Table 1 : Variations of the solutions. The linear perturbation equations to be solved and the basic conditions of resistivity $\epsilon$ and viscosity $N$ are summarized for each case.

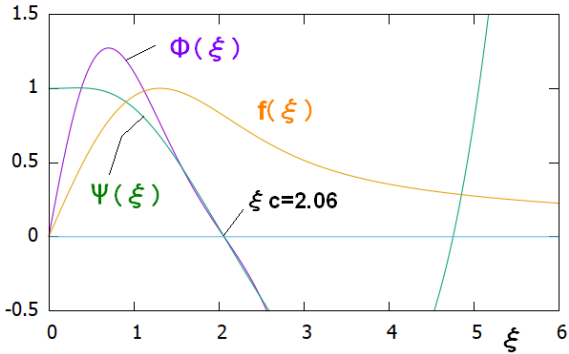

Fig. 1 : Zero-crossing solution of Case 1 for $\kappa = 1$, $\epsilon = 0.1$, $N = 0$, $\Phi'(0) = 3.0802$, $\lambda = 0.13$, and $\xi_c = 2.06$.

### 2.2.1 Case 1a: $\lambda = 0$ solution

It is worth studying the case of $\lambda = 0$, which gives the critical condition of the instability. Taking $\lambda = 0$, Eqs. (8) and (9) are changed as follows:

$$\Psi'' = (\kappa^2 \epsilon^2 + f''(\xi)/f(\xi))\Psi \qquad (10)$$

$$f''(\xi)\Psi = -\kappa f(\xi)^2 \Phi. \qquad (11)$$

By solving Eq.(10) as an IVP in $0 < \xi < \xi_c$, $\Psi$ can be obtained. Note that Eq.(7) gives $f(0) = 0$ but $f''(0)/f(0) = -2$ in the $\xi = 0$ limit. Then, $\Phi$ is directly obtained from Eq. (11) in $\xi > 0$ but diverges at $\xi = 0$. The divergence has been reported in the Appendix of Shimizu's paper [36]. Then, zero-crossing solutions are obtained. Eq.(10) means that $\Psi$ depends only on $\kappa\epsilon$. In other words, $\Psi$ does not depend separately on $\kappa$ and $\epsilon$. This $\kappa\epsilon$ dependence for $\lambda = 0$ is inapplicable for $\Phi$ but is applied for $\kappa\Phi$ and is also maintained in Cases 2a, 3a, 4a, and 5a shown below.

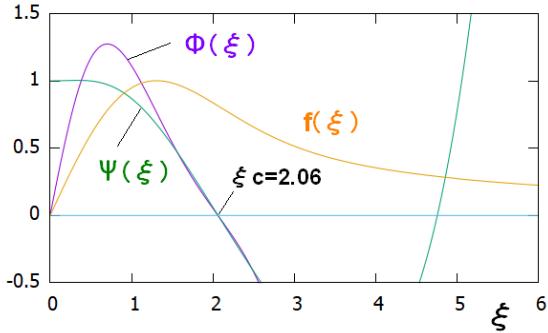

Fig. 2 : Zero-crossing solution of Case 2 for $\kappa = 1$, $\epsilon = 0.1$ in $\xi < 1.307$ (= 0 in $\xi > 1.307$), $N = 0$, $\Phi'(0) = 3.0804$, $\lambda = 0.13$, and $\xi_c = 2.06$.

### 2.2.2 Case 1b: $\epsilon = 0$ solution

This is a special case where $\epsilon = 0$ is taken in Eqs. (8) and (9). The result in this case may be called ideal tearing instability. In this case, zero-crossing solutions are found [36].

## 2.3 Case 2: Non-Uniform Resistivity and No Viscosity

In this case, the resistivity $\epsilon$ is uniform in $\xi \leq 1.307$, i.e., the inner region of the current sheet but is zero in $\xi > 1.307$, i.e., the outer region. Hence, this case is close to the traditional setup of the original LSC theory [8]. The viscosity $N$ is uniformly zero, i.e., inviscid, in $0 < \xi < \xi_c$. In this case, in $\xi \leq 1.307$, Eqs. (8) and (9) are solved. Then, in $\xi > 1.307$, since $\epsilon = 0$ is set in Eqs. (8) and (9), the next equations are solved.

$$0 = \lambda\Phi'' + f(\xi)\Psi'' - f''(\xi)\Psi \tag{12}$$

$$\Psi'' = \kappa\lambda\Psi - \kappa f(\xi)\Phi. \tag{13}$$

Eventually, Eqs. (8), (9), (12), and (13) can be numerically solved as IVP. Then, zero-crossing solutions are found. Fig. 2 shows the typical profile of the solutions, which are obtained for $\kappa = 1$, $\epsilon = 0.1$ (for $\epsilon \leq 1.307$), $N = 0$, $\lambda = 0.13$, and $\Phi'(0) = 3.0804$. This is a zero-crossing solution at $\xi_c = 2.06$. Fig. 2 is almost the same as Fig. 1. However, as shown later, we cannot say that Cases 1 and 2 are exactly the same. In contrast to Fig. 1, since $\epsilon$ discontinuously changes from a non-zero value to zero at $\xi = 1.307$, $\Phi''$ and $\Psi''$ generally have discontinuities at the point, but the discontinuity is invisible in Fig. 2.

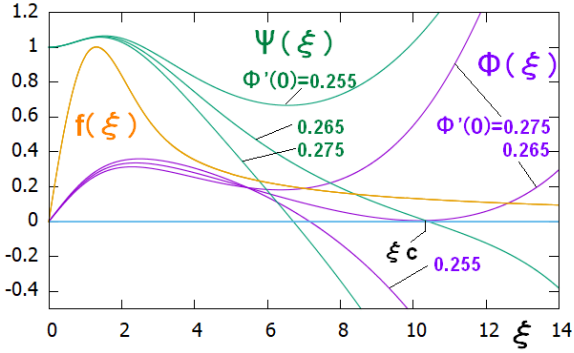

Fig. 3 : Zero-contact solution of Case 3 for $\kappa = 1$, $\epsilon = 0.1$, $N = 0.05$, $\Phi'(0) = 0.265$, (0.255 and 0.275), $\Phi'''(0) = -0.2253415$, $\lambda = 0.13$, and $\xi_c = 10.28$.

### 2.3.1  Case 2a: $\lambda = 0$ solution

To study the case of $\lambda = 0$, we take the $\lambda = 0$ limit. In $\xi \leq 1.307$, Eqs. (10) and (11) are solved. Then, in $\xi > 1.307$, setting $\epsilon = 0$, Eq. (10) is changed as follows:

$$\Psi'' = (f''(\xi)/f(\xi))\Psi \tag{14}$$

Hence, in $\xi > 1.307$, Eqs. (11) and (14) are solved. Then, zero-crossing solutions are found.

## 2.4  Case 3: Uniform Resistivity and Uniform Viscosity

This case is the most highlighted case in this paper. In contrast to Cases 1 and 2, the viscosity effect is introduced. Both the resistivity $\epsilon$ and viscosity $N$ are uniform in $0 < \xi < \xi_c$. The equations to be solved are as follows (the derivation is found in Appendix):

$$N\Phi'''' = \kappa\epsilon^2((\lambda + 2\kappa N)\Phi'' - (\lambda + \kappa N)\kappa^2\epsilon^2\Phi + f(\xi)(\Psi'' - \kappa^2\epsilon^2\Psi) - f''(\xi)\Psi) \tag{15}$$

$$\Psi'' - \kappa^2\epsilon^2\Psi = \kappa\lambda\Psi - \kappa f(\xi)\Phi \tag{16}$$

Eq. (16) is exactly the same as Eq. (9). In contrast to Cases 1 and 2, the fourth-order derivative of $\Phi$ appears because of the introduction of the viscosity. This results in the number of initial values in the IVP increasing, and hence, the number of zero-crossing solutions found at a specified $\xi_c$ becomes infinite. To uniquely specify one from those zero-crossing solutions, we add another upstream open boundary condition $\Phi' = 0$ at $\xi_c$. We define it as a zero-contact solution. In fact, such zero-contact solutions are found. Fig. 3 shows the typical profiles of the solutions, which are obtained for $\kappa = 1$, $\epsilon = 0.1$, $N = 0.05$, $\lambda = 0.13$, $\Phi'(0) = (0.255, 0.265, 0.275)$, and $\Phi'''(0) = -0.2253415$. The profiles of $\Phi$ and $\Psi$ for $\Phi'(0) = 0.265$ are zero-contact solutions at $\xi_c = 10.28$.

### 2.4.1 Case 3a: $\lambda = 0$ solution

For $\lambda = 0$, Eqs. (15) and (16) are changed as follows:

$$N\Phi'''' = \kappa\epsilon^2(2\kappa N\Phi'' - \kappa^3\epsilon^2 N\Phi - \kappa f(\xi)^2\Phi - f''(\xi)\Psi) \tag{17}$$

$$\Psi'' - \kappa^2\epsilon^2\Psi = -\kappa f(\xi)\Phi. \tag{18}$$

By solving these equations, the zero-contact solutions can be obtained.

### 2.4.2 Case 3b: $N = +\infty$ solution

This case is an extreme case. Taking the $N = +\infty$ limit, Eq. (15) is changed as follows.

$$\Phi'''' = 2\kappa^2\epsilon^2\Phi'' - \kappa^4\epsilon^4\Phi \tag{19}$$

In addition to this equation, Eq. (16) is solved. In this case, zero-contact solutions are not found. This suggests that the tearing instability in Case 3 does not occur for $N = +\infty$. In other words, it suggests that sufficiently large viscosity can steadily stabilize the tearing instability, as extensively discussed in Chapter 5.

## 2.5 Case 4: Uniform Resistivity and Non-Uniform Viscosity with continuous $\Phi''$ at $\xi = 1.307$

Cases 1 and 2 can be compared to explore how the non-uniformity of the resistivity affects the instability. Similarly, the non-uniformity of the viscosity may be studied. In this case, Eqs. (15) and (16) are solved in $\xi \leq 1.307$. Then, by setting $N = 0$ in those equations, Eqs. (8) and (9) are solved in $\xi > 1.307$. The resistivity $\epsilon$ is uniform in $0 < \xi < \xi_c$. To maintain the continuity of $\Phi''$ at $\xi = 1.307$, Eqs. (8) and (15) must be simultaneously satisfied at $\xi = 1.307$. At the time, Eq. (19) must be satisfied at $\xi = 1.307 - 0$, which is in the vicinity of $\xi = 1.307$ on the $\xi < 1.307$ side. Then, zero-crossing solutions are found. Fig. 4 shows the typical profile of the solutions for $\kappa = 1$, $\epsilon = 0.1$, $N = 0.05$ (for $\xi < 1.307$), $\lambda = 0.13$, $\Phi'(0) = 0.04428$, and $\Phi'''(0) = -1.73538$, which is a zero-crossing solution at $\xi_c = 2.806$.

### 2.5.1 Case 4a: $\lambda = 0$ solution

For $\lambda = 0$, Eqs. (17) and (18) are solved in $\xi \leq 1.307$. In $\xi > 1.307$, because of $N = 0$ and $\lambda = 0$, Eqs. (10) and (11) are solved. Since Eqs. (10), (11), (17) and (18) must be simultaneously satisfied at $\xi = 1.307$, Eq. (11) must be satisfied at $\xi = 1.307 - 0$, resulting in the continuity of $\Phi$ being maintained at $\xi = 1.307$. In addition, since $\Phi$ in $\xi > 1.307$ is directly obtained from $\Psi$ through Eq. (11), $\Phi'$ is not necessarily continuous at $\xi = 1.307$. To maintain the continuity of $\Phi'$ at this point, the next condition must be satisfied at $\xi = \xi_0 = 1.307$.

$$\Phi'(\xi_0 - 0) = \Phi'(\xi_0 + 0), \tag{20}$$

Eventually, by solving Eqs. (17), (18), (10) and (11) as IVP with additional conditions of Eqs. (11) and (20) at $\xi = 1.307$, zero-crossing solutions are obtained.

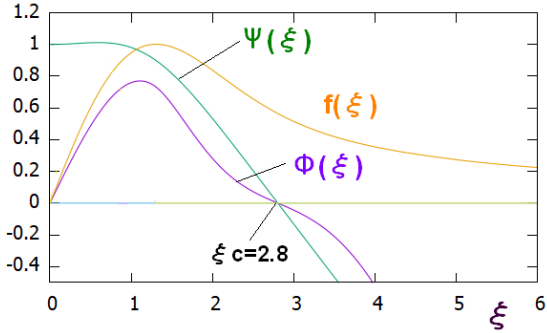

Fig. 4 : Zero-crossing solution of Case 4 for $\kappa = 1$, $\epsilon = 0.1$, $N = 0.05$ in $\xi < 1.307$ ($N = 0$ in $\xi > 1.307$), $\Phi'(0) = 0.04428$, $\Phi'''(0) = -1.73538$, $\lambda = 0.13$, and $\xi_c = 2.8$.

### 2.5.2 Case 4b: $N = \infty$ solution

This case is an extreme case. By setting the limit as $N = \infty$, Eq. (15) is changed to Eq. (19). Hence, in $\xi \leq 1.307$, Eqs. (16) and (19) are solved. In $\xi > 1.307$, since $N = 0$ is set, Eqs. (8) and (9) are solved. To maintain the continuity of $\Phi''$ at $\xi = 1.307$, Eq. (8) must be satisfied at $\xi = 1.307 - 0$. Then, zero-crossing solutions are obtained.

### 2.5.3 Case 4c: $N = \infty$ and $\lambda = 0$ solution

This case is more extreme than Cases 4a and 4b. In $\xi \leq 1.307$, because of $\lambda = 0$, Eqs. (18) and (19) are solved. In $\xi > 1.307$, because of $N = 0$ and $\lambda = 0$, Eqs. (10) and (11) are solved. Eq. (11) must be satisfied at $\xi = 1.307 - 0$, to maintain the continuity of $\Phi$, i.e., $\Phi(1.307 - 0) = \Phi(1.307 + 0)$. In addition, to uniquely specify one of those zero-crossing solutions, we maintain the continuity of $\Phi'$ at $\xi = 1.307$. To do so, Eq. (20) must be satisfied at $\xi = 1.307$. Then, zero-crossing solutions are obtained.

## 2.6 Case 5: Uniform Resistivity and Non-Uniform Viscosity with Discontinuous $\Phi''$ at $\xi = 1.307$

Case 4 is suitable for comparison with Case 1 because the upstream boundary condition at $\xi_c$ is the same, i.e., where zero-crossing solutions are obtained. However, Case 4 is not suitable for comparison with Case 3 because the upstream boundary condition is different. To compare with Case 3, we study when Eqs. (15), (16), (8), and (9) are solved without the connecting condition, i.e., Eq. (19), at $\xi = 1.307$. In this case, zero-contact solutions can be found. Fig. 5 shows the typical profile of the solutions for $\kappa = 1$, $\epsilon = 0.1$, $N = 0.05$ (for $\xi < 1.307$), $\lambda = 0.13$, $\Phi'(0) = 0.69407987$, and $\Phi'''(0) = -0.7857$, which is a zero-contact solution at $\xi_c = 3.5$. In this case, $\Phi''$ is discontinuous at $\xi = 1.307$. However, $\Phi'$ and $\Phi$ remain continuous at this point. Hence, the constraint of the differential discontinuity is looser than Case 4, instead of more rigid upstream boundary condition. Considering

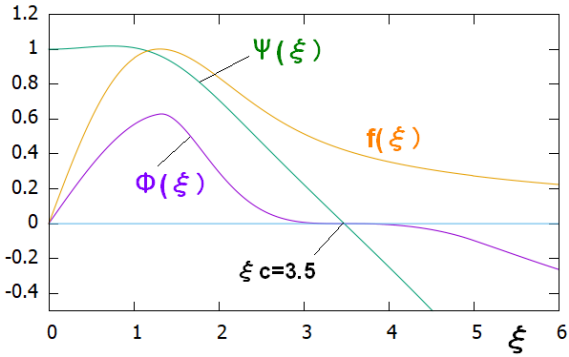

Fig. 5 : Zero-contact solution of Case 5 for $\kappa = 1$, $\epsilon = 0.1$, $N = 0.05$ in $\xi < 1.307$ (= 0 in $\xi > 1.307$), $\Phi'(0) = 0.69407987$, $\Phi'''(0) = -0.785715$, $\lambda = 0.13$, and $\xi_c = 3.5$.

Eqs. (8) and (9) with $\Phi = \Psi = 0$, $\Phi'' = \Psi'' = 0$ is satisfied at $\xi_c$. This means that $\xi_c$ is the inflection point, as shown in Fig. 5.

### 2.6.1 Case 5a: $\lambda = 0$ solution

In this case, Eqs. (17) and (18) are solved in $\xi \le 1.307$. Then, Eqs. (11) and (18) are solved in $1.307 < \xi$. In addition, to maintain the continuity of $\Phi$ at $\xi = 1.307$, $\Phi(1.307 - 0) = \Phi(1.307 + 0)$ is applied. In this case, the continuity of $\Phi'$ is not maintained at $\xi = 1.307$. Then, zero-contact solutions for $\xi_c = 2.12$ are found regardless of $\kappa$, $\epsilon$, and $N$.

## 3 Methods of the Numerical Study

### 3.1 Initial Value Problem (IVP)

Traditionally, the equations shown in Chapter 2 are solved as eigenvalue problems (EVPs), where $\Phi$ and $\Psi$ are obtained in $0 < \xi < \infty$. However, in this paper, those equations are solved as the IVPs of $0 < \xi < \xi_c$, where $\xi_c$ is a finite value, i.e., $\xi_c < +\infty$. With increasing $\xi_c$, those solutions in $0 < \xi < +\infty$ may be indirectly deduced. In IVPs, a set of $\kappa$, $\epsilon$, $N$, and $\lambda$ is firstly given. Then, a set of the initial values given at $\xi = 0$ consists of $\Phi(0)$, $\Psi(0)$, $\Phi'(0)$, $\Phi''(0)$, $\Phi'''(0)$, and $\Psi'(0)$. Notably, $\Phi''(0)$ and $\Phi'''(0)$ are not needed in Cases 1 and 2. Then, through the parameter survey, zero-crossing solutions or zero-contact solutions are numerically found by adjusting those initial values in the manners shown below.

We consider the symmetric configuration of tearing instability at $\xi = 0$. In that case, since $\Phi$ and $\Psi$ are odd and even functions, respectively, $\Phi(0) = \Phi''(0) = 0$ and $\Psi'(0) = 0$ are set. In addition, $\Psi(0) = 1$ can be set without the lack of generality of solutions. Eventually, $\Phi'(0)$ and $\Phi'''(0)$ are the control parameters to uniquely determine a solution in the IVP.

To numerically solve $\Phi$ and $\Psi$ as IVPs, the primitive forward Euler method is employed throughout this paper. The numerical resolution is mainly set in $\Delta\xi = 0.001$. Some of the following results are extremely sensitive to changing those initial values. For much higher numerical resolutions, i.e., $0.001 > \Delta\xi$, the convergence test of the numerical results has been carefully checked.

## 3.2   Obtaining the Solutions

### 3.2.1   Parameter Survey of IVPs

In this chapter, we explain how to find the zero-contact solutions in Case 3. The application to the other cases is similar or much easier. The actual numerical results for Cases 1-5 are reported in the next chapter. In Case 3, the IVP starts from $\Phi(0) = \Phi''(0) = 0$, $\Psi(0) = 1$ and $\Psi'(0) = 0$, where $\kappa$, $\epsilon$, and $N$ are initially given. In addition, $\lambda$, $\Phi'(0)$ and $\Phi'''(0)$ are the adjustable control parameters. Once these three parameters are given, a set of $\Phi(\xi)$ and $\Psi(\xi)$ is uniquely determined by solving the IVP. By adjusting $\lambda$, $\Phi'(0)$ and $\Phi'''(0)$, the IVP can be repeatedly solved. Then, we can find a set of $\Phi$ and $\Psi$ that satisfy $\Phi = \Psi = \Phi' = 0$ at $\xi_c$, if it exists. That is, a zero-contact solution can be obtained. Since this is a numerical study of IVPs, we cannot confirm the uniqueness of the solution for a set of $\kappa$, $\epsilon$, and $N$. At this point, our interest is to find the solution that provides the highest growth rate but it is unclear whether the obtained solution has the highest rate. However, if the solution is found for $\lambda > 0$, tearing instability can, at least, occur at the $\lambda$ rate.

### 3.2.2   Profiles of $\Phi$ and $\Psi$

Fig. 3 shows how the numerically obtained profiles of $\Phi$ and $\Psi$ change, depending on the value of $\Phi'(0)$, where $\kappa = 1$, $\epsilon = 0.1$, $N = 0.05$, $\lambda = 0.13$, and $\Phi'''(0) = -0.2253415$. The purple lines indicate the profiles of $\Phi$. The green lines indicate the profiles of $\Psi$. The equilibrium magnetic field function $f(\xi)$ is indicated by the orange line, which is fixed throughout this paper. As $\Phi'(0)$ increases from 0.255 to 0.275, the profile of $\Phi$ tends to gradually shift upward, and the profile of $\Psi$ tends to gradually shift downward. For $\Phi'(0) = 0.255$, the zero-crossing point of $\Phi$, i.e., $\Phi = 0$, is observed around $\xi = 7$. However, the zero-crossing point of $\Psi$, i.e., $\Psi = 0$, is not observed in this figure. Conversely, in the case of $\Phi'(0) = 0.275$, the zero-crossing point of $\Psi$ is observed at approximately $\xi = 6.5$. However, the zero-crossing point of $\Phi$ is not observed in this figure. Finally, in the case of $\Phi'(0) = 0.265$, the zero-crossing points of $\Phi$ and $\Psi$ almost coincide around $\xi_c = 10.3$. In addition, it seems that $\Phi' = 0$ is established at this point. Hence, $\Phi$ and $\Psi$ for $\Phi'(0) = 0.265$ seem to be a zero-contact solution. However, Fig. 3 does not indicate that $\Phi$ and $\Psi$ for $\Phi'(0) = 0.265$ are exactly the zero-contact solution. More exact observations are required to confirm this hypothesis. To do so, the next two strategies, i.e., status map and trajectory of the crossing points, are proposed.

### 3.2.3   Status Map

To rigorously specify the existence of the zero-contact solution, we focus on the movements of the zero-crossing points of $\Phi$ and $\Psi$ and local minimum point of $\Phi$. we define those locations as $\xi_1$, $\xi_2$ and $\xi_3$, respectively. Table 2 shows the classification list of solutions for $\xi_1$, $\xi_2$, and $\xi_3$, which are numerically detected in the parameter survey of IVPs. Then, the zero-contact solution satisfies $\xi_1 = \xi_2 = \xi_3$, which is $\xi_c$.

   Fig. 6(a) shows a typical status map of the numerically obtained solutions, where the types of solutions are classified by the colors listed in Table 2. The vertical axis indicates the change in $\Phi'(0)$, and the horizontal axis indicates the change in $\Phi'''(0)$, where $\kappa = 1$, $\epsilon = 0.1$, $N = 0.05$, and $\lambda = 0.13$ are fixed. Fig. 6(a) is drawn by stepwisely changing $\Phi'(0)$ and $\Phi'''(0)$ in each axis range, where the IVPs are repeatedly solved.

| Condition(Color) | Cross Pt. of $\Phi = 0$, | Cross Pt. of $\Psi = 0$, | Local Minimum of $\Phi' = 0$, | Order of Locations |
|---|---|---|---|---|
| Condition 1 (Pink) | $\xi_1 > 0$ | not detected | not detected | |
| Condition 1 (Pink) | $\xi_1 > 0$ | not detected | $\xi_3 > 0$ | |
| Condition 2 (Yellow) | $\xi_1 > 0$ | $\xi_2 > 0$ | not detected | $\xi_1 < \xi_2$ |
| Condition 2 (Yellow) | $\xi_1 > 0$ | $\xi_2 > 0$ | $\xi_3 > 0$ | $\xi_1 < \xi_2 < \xi_3$ |
| Condition 3 (Cyan) | $\xi_1 > 0$ | $\xi_2 > 0$ | not detected | $\xi_2 \leq \xi_1$ |
| Condition 4 (Light-Green) | $\xi_1 > 0$ | $\xi_2 > 0$ | $\xi_3 > 0$ | $\xi_1 < \xi_3 \leq \xi_2$ |
| Condition 5 (Dark-Green) | $\xi_1 > 0$ | $\xi_2 > 0$ | $\xi_3 > 0$ | $\xi_2 \leq \xi_1 < \xi_3$ |
| Condition 6 (Black) | not detected | $\xi_2 > 0$ | not detected | |
| Condition 7 (Orange) | not detected | not detected | $\xi_3 > 0$ | |
| Condition 8 (Red) | not detected | $\xi_2 > 0$ | $\xi_3 > 0$ | $\xi_3 \leq \xi_2$ |
| Condition 9 (Blue) | not detected | $\xi_2 > 0$ | $\xi_3 > 0$ | $\xi_2 < \xi_3$ |

Table 2 : Types of solutions in the status maps of Figs.6(a) and (b). Each type is related to the colors in the map. Cyan and black, which are listed here, are not observed in this paper because the conditions are not matched.

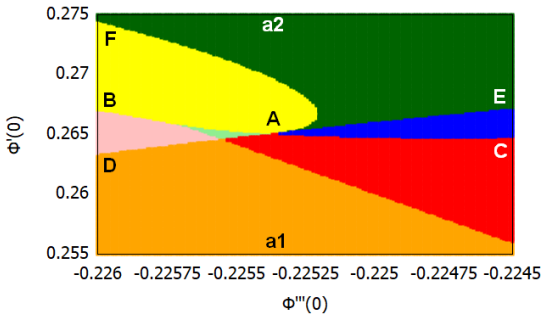

Fig. 6(**a**): Status map of Case 3 in the wide range, i.e., $0.255 < \Phi'(0) < 0.275$ and $-0.2260 < \Phi'''(0) < -0.2245$, for $\kappa = 1$, $\epsilon = 0.1$, $N = 0.05$, and $\lambda = 0.13$.

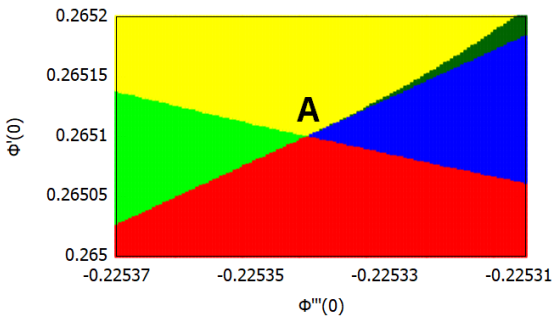

Fig. 6(**b**): Status map of Case 3 in the narrow range, i.e., $0.2650 < \Phi'(0) < 0.2652$ and $-0.22537 < \Phi'''(0) < -0.22531$, for $\kappa = 1$, $\epsilon = 0.1$, $N = 0.05$, and $\lambda = 0.13$.

The case of $\Phi'(0) = 0.265$ in Fig. 3 corresponds to Point A in Fig. 6(a). Then, the cases of $\Phi'(0) = 0.255$ and $0.275$ are respectively located at Points a1 and a2 in Fig. 6(a). As referred in Table 2, Line B-A-C indicates the boundary line of whether $\xi_2 < \xi_3$ or not. Line D-A-E indicates the boundary line of whether $\Phi(\xi_3) > 0$ or not. Line D-A-F indicates the boundary line of whether $\xi_1 < \xi_2$ or not. Since the zero-contact solution should be exactly located on these three lines, Point A exactly indicates the location of the zero-contact solution.

In this discussion, we assumed that the profiles of $\Phi$ and $\Psi$ smoothly change for every parameter change, such as $\lambda$, $\Phi'(0)$ and $\Phi'''(0)$. Inversely speaking, if those profiles discontinuously change, we cannot say that the zero-contact solution exists at Point A. This assumption can be a serious concern. At this point, as shown in Fig. 6(b), the enlargement of the status map around Point A indicates that the profiles smoothly change. Fig. 6(b) shows the status map near Point A. The figure format is the same as that of Fig. 6(a). In comparison with Fig. 6(a), the parameter survey range of $\Phi'(0)$ and $\Phi'''(0)$ in Fig. 6(b) is much narrower, and Point A can still be observed on the three lines. Hence, we conclude that a zero-contact solution exists at Point A.

### 3.2.4   Trajectory of the Crossing Points

In this section, another method is proposed to steadily find zero-contact solutions. Certainly, Fig. 6 can be used to conveniently find zero-contact solutions. However, this method is not perfect. For example, when those three boundary lines observed in the status maps almost merge to a line around A, it is impossible to exactly specify the location of Point A. In fact, such situations often occur. As another problem, the zero-contact solutions for the second and third zero-crossing points can appear, as later shown in Fig. 9(c). In such cases, Table 2 and the status map must be largely improved. Rather, Fig. 7 is more helpful to find the zero-contact solutions.

Fig. 7 shows how the zero-crossing points of $\Phi$ and $\Psi$ move in the $\xi$ space, depending on the value of $\Phi'(0)$ for a fixed $\Phi'''(0)$. Figs. 7(a), (b), and (c) are plotted for $\Phi'''(0) = -0.22532$, $-0.22537$, and $-0.2253415$, respectively. In these figures, the row of circles indicates the trajectory of the zero-crossing points of $\Psi$, i.e., where $\Psi = 0$ is satisfied. Similarly, the row of the cross-points indicates that of $\Phi$, i.e., where $\Phi = 0$ is satisfied.

Point A shown in Figs. 7(a), (b), and (c) indicates the local minimum point of $\Phi$, i.e., where $\Phi' = 0$ is satisfied. Point B indicates the zero-crossing solution, i.e., when $\Phi = \Psi = 0$ is satisfied. In Fig. 7(a), Point A is located on the upper side of Point B. In Fig. 7(b), Point A is located on the lower side of Point B. As $\Phi'''(0)$ changes from Fig. 7(a) to (b), we assume that the trajectories shown in these figures smoothly change. In that case, a zero-contact solution should be located between Figs. 7(a) and (b). Fig. 7(c) is plotted for the value of $\Phi'''(0)$ located between Figs. 7(a) and (b). In Fig. 7(c), since Points A and B coincide, they are considered a zero-contact solution. Eventually, we found the zero-contact solution for $\kappa = 1$, $\epsilon = 0.1$, $N = 0.05$, $\lambda = 0.13$, $\Phi'(0) = 0.2650997$, and $\Phi'''(0) = -0.2253415$, which coincides with Point A in Fig. 6(b).

## 4   Numerical results

### 4.1   Overview of the relation of $\lambda$ and $\xi_c$

By using the status maps and trajectories of the crossing points, a set of $\kappa$, $\epsilon$, $N$, $\lambda$, and $\Phi'(0)$, $\Phi'''(0)$ yields a zero-contact solution for a $\xi_c$ value in Cases 3 and 5. In addition, it yields a zero-crossing solution for a $\xi_c$ value in Cases 1, 2, and 4. Then, we can know the relation of $\lambda$ and $\xi_c$ in each case.

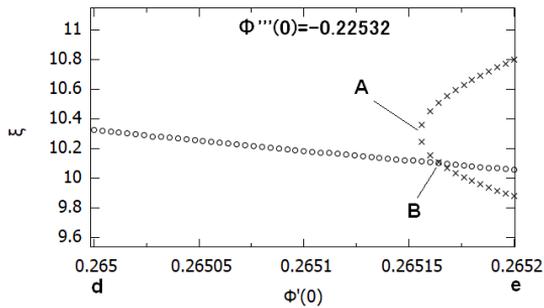

Fig. 7(**a**): Trajectories of zero-crossing points of $\Phi$ and $\Psi$ in the range of $0.2650 < \Phi'(0) < 0.2652$ for $\kappa = 1$, $\epsilon = 0.1$, $N = 0.05$, $\lambda = 0.13$, and $\Phi''(0) = -0.22532$.

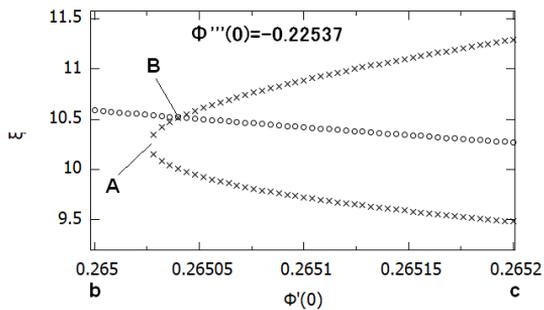

Fig. 7(**b**): The case of $\Phi'''(0) = -0.22537$. The others are the same as (a).

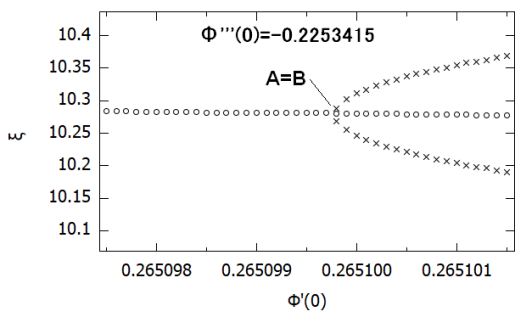

Fig. 7(**c**): The case of $\Phi'''(0) = -0.2253415$. The others are the same as (a).

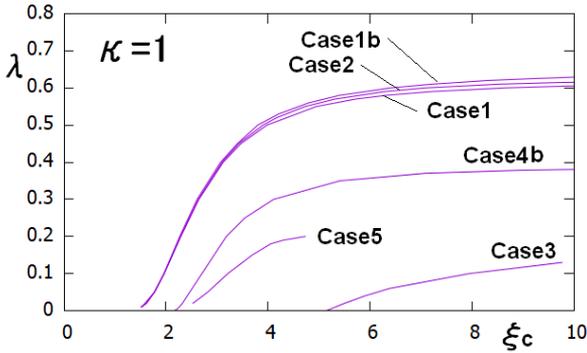

Fig. 8 : Linear growth rate $\lambda$ with respect to $\xi_c$ for $\kappa = 1$ and $\epsilon = 0.1$. Cases 1 and 2 are plotted for $N = 0$. Cases 3 and 5 are plotted for $N = 0.05$. Case 4b is plotted for $N = +\infty$.

Fig. 8 shows how $\lambda$ depends on $\xi_c$, where $\kappa = 1$ and $\epsilon = 0.1$, except for Case 1b, where $\epsilon = 0$. The lines of Cases 1, 1b and 2 are plotted for $N = 0$, i.e., inviscid. The lines of Cases 3 and 5 are plotted for $N = 0.05$. The line of Case 4b is for $N = +\infty$. First, it is remarkable that every case plotted in Fig. 8 shows that as $\xi_c$ increases, $\lambda$ monotonically increases. This monotonical increase is commonly observed through this paper, i.e., any case and any condition. Second, $\lambda$ of Case 1b takes the highest $\lambda$ value in the entire range of $\xi_c$, which is the case of $\epsilon = N = 0$, i.e., the ideal-MHD limit. The existence of the solutions in the ideal-MHD limit has been reported [35,36]. Subsequently, the order of Cases $2 > 1 > 4b > 5 > 3$ is observed for the height of $\lambda$. The lines of Cases 3 and 5 disappear in $\xi_c > 9.8$ and $\xi_c > 4.9$, respectively. These disappearances originate from the lack of numerical precision in the IVPs In general, it is harder to find solutions for larger $\xi_c$. Third, $\lambda$ of Cases 1, 1b, and 2 almost coincide. In fact, the difference is below approximately a few %. Regarding this point, Figs. 1 and 2 are almost the same. This suggests that $\lambda$ obtained in the modified LSC theory is close to that of the original LSC theory. However, as shown in Fig. 9(a), the difference between Cases 1 and 2 becomes larger for larger $\kappa\epsilon$. Fourth, the line of Case 4b is plotted in $N = +\infty$, as discussed in Fig. 13(b), later.

## 4.2 Cases 1, 2, and 3

Fig. 9(a) shows the relation of $\kappa\epsilon$ and $\xi_c$ for $\lambda = 0$ of Cases 1a, 2a, and 3a, which is the critical condition of the tearing instability, i.e., beyond which instability occurs. In fact, above these lines, instability occurs, and stability occurs below. Because, the solutions of $\lambda > 0$ are found only above those lines. What is shown in this figure coincides with the foot points, i.e., $\lambda = 0$, of each line shown in Fig. 8. In this figure, since $\epsilon = 0.1$ is fixed, the horizontal scale of this figure directly indicates that $0.1\kappa$. At this point, as mentioned in the preceding chapter, the solutions of $\lambda = 0$ depend on $\kappa\epsilon$ and does not separately depend on $\kappa$ and $\epsilon$. Hence, this figure is applicable in $0 < \epsilon < +\infty$.

Lines A and B in Fig. 9(a) are for Cases 1a and 2a and stay around $\kappa\epsilon = 1.1$ and 1.3, respectively, in $\xi_c > 5$. Hence, the right region of those lines is stable, and the left is unstable. Since Case 2 is close to FKR theory [2], $\kappa\epsilon = 1.3$ on Line B may correspond to the critical condition of $\Delta' = 0$ in the theory. In fact, the value of $\kappa\epsilon = 1.3$ is roughly close to the predicted $\alpha_c = 1.0$ and 0.64, respectively, in Eqs. (28) and (30) in FKR theory. Lines c1-d1, c2-d2, c3-d3, and c4-d4 are obtained for Case 3a, which represent $N = 0.0005$, 0.002, 0.05 and 0.1, respectively. In contrast to Cases 1a and 2a, when $\kappa\epsilon$ approaches zero, these lines of Case 3a diverge toward $\xi_c = +\infty$. This means that the stable region appears in the low $\kappa\epsilon$ region by the introduction of viscosity. This important characteristic is extensively discussed later.

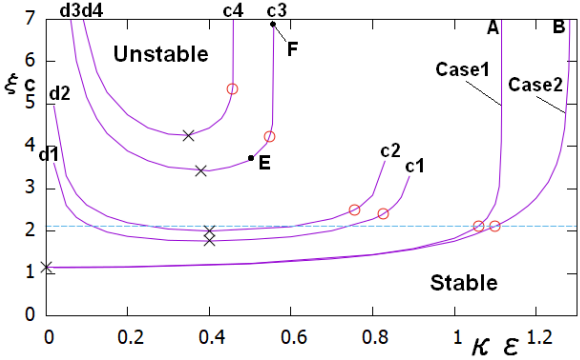

Fig. 9(**a**): Critical conditions of Cases 1, 2, and 3. Cases 1 and 2 are plotted for $N = 0$. Case 3 is plotted for 0.0005(c1-d1), 0.002(c2-d2), 0.05(c3-d3), and $N = 0.1$(c4-d4). The red circles indicate the change point of the solution types. The black cross points indicate the minimum point of $\xi_c$, beyond which the unstable range appears.

The six red circles in Fig. 9(a) indicate when the crossing point of $\Phi = 0$ switches from the first point to the second, as shown in the change from Figs. 9(b) to (c). Figs. 9(b) and (c) are zero-contact solutions at $\lambda = 0$, i.e., Case 3a. The solution shown in Fig. 9(b) is located at Point E in Fig. 9(a), which is to the left of the red circle. This solution takes the first zero-crossing point of $\Phi = \Psi = 0$ at $\xi_c = 3.65$. Moreover, the solution shown in Fig. 9(c) is located at Point F in Fig. 9(a), which is to the right (exactly upside) of the red circle. As shown in Fig. 9(c), this solution takes the first zero-crossing point of $\Phi = 0$ at $\xi = 3.069$, where $\Psi \neq 0$. The zero-crossing point of $\Psi = 0$ is located at $\xi_c = 6.91$, where $\Phi = 0$. Then, the zero-contact solution is established at $\xi_c = 6.91$. As shown in Fig. 9(a), the zero-contact point $\xi_c$ tends to change from the first zero-crossing point to the second zero-crossing point as $\kappa\epsilon$ increases. In fact, every line of Fig. 9(a) has a circle.

Fig. 10(a) shows how $\lambda$ changes for $N$ for $\epsilon = 0.1$ and $\xi_c = 3.6$ for Cases 1 and 3. In this figure, $\lambda$ at $N = 0$ (i.e., on the vertical axis) is obtained in Case 1. Then, $N$ at $\lambda = 0$ (i.e., horizontal axis) is obtained in Case 3a, i.e., the $\lambda = 0$ limit. The $N$ value at $\lambda = 0$ is also observed in Fig. 9(a). As shown in this figure, $\lambda$ always takes the maximum value at $N = 0$. Then, as $N$ increases, $\lambda$ monotonically decreases to zero.

Fig. 10(b) shows the relation of $\lambda$ and $\xi_c$ for Cases 1 and 3. The figure format is the same as that of Fig. 8. The two lines for $N = 0.05$ are much lower than each corresponding line for $N = 0$. Hence, the viscosity effect tends to steadily slow the tearing instability. Then, focusing only on the two lines of $N = 0$, the line of $\epsilon = 0.1$ is found to be higher than that of $\epsilon = 0.2$. This means that, as the resistivity $\epsilon$ is close to zero, the linear growth rate $\lambda$ is higher. Inversely, focusing on the two lines of $N = 0.05$, the line of $\epsilon = 0.1$ is lower than that of $\epsilon = 0.2$. This inversion will be discussed in Fig. 10(d), later.

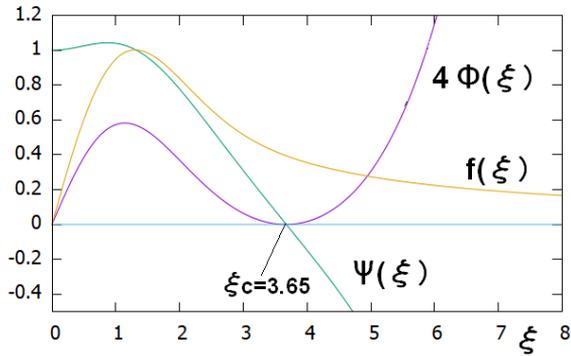

Fig. 9(**b**): Zero-contact solution of Case 3a for $\kappa = 5.0$, $\epsilon = 0.1$, $N = 0.05$, $\lambda = 0$, $\Phi'(0) = 0.212044$, and $\Phi'''(0) = -0.512955$.

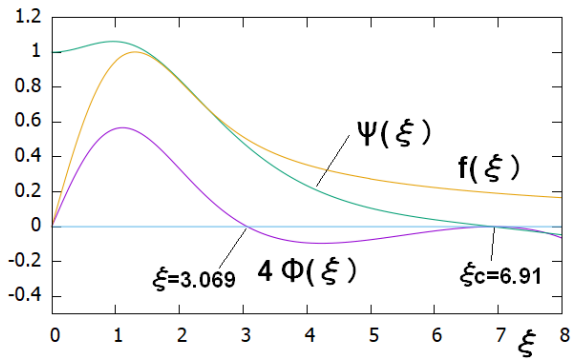

Fig. 9(**c**): Zero-contact solution of Case 3a for $\kappa = 5.58$, $\epsilon = 0.1$, $N = 0.05$, $\lambda = 0$, $\Phi'(0) = 0.212156$, and $\Phi'''(0) = -0.538406$.

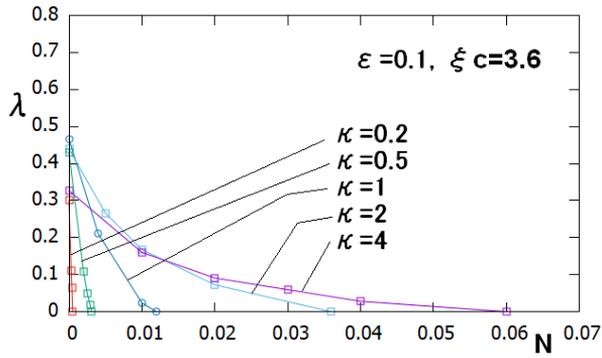

Fig. 10(**a**): Linear growth rate $\lambda$ with respect to $N$ in Case 3 for $\kappa = 0.2, 0.5, 1, 2, 4$ and $\epsilon = 0.1$, where $\lambda$ at $N = 0$ is Case 1.

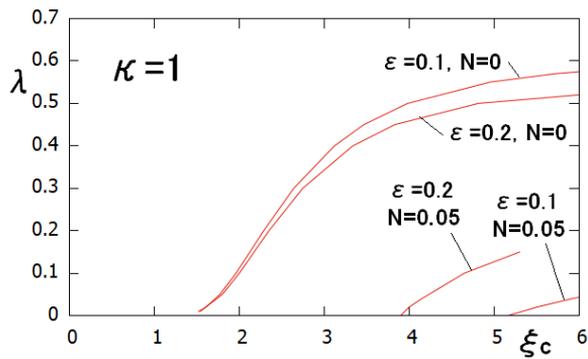

Fig. 10(**b**)Linear growth rate $\lambda$ with respect to $\xi_c$ in Case 3 for $\kappa = 1$, $\epsilon = 0.1, 0.2$ and $N = 0, 0.05$.

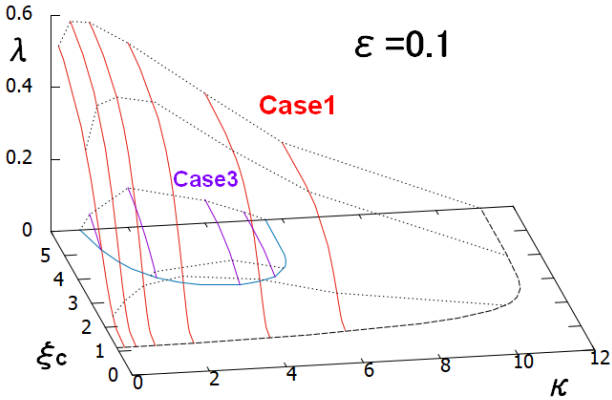

Fig. 10(**c**): 3D image of linear growth rate $\lambda$ with respect to $\xi_c$ and $\kappa$ in Case 1 for $\epsilon = 0.1$ and $N = 0$ and Case 3 for $\epsilon = 0.1$ and $N = 0.05$.

Fig. 10(c) shows the 3D image of Figs. 9(a) and 10(b), where the bottom plane of $\kappa - \xi_c$ is drawn for Cases 1a and 3a (Line c3-d3) in Fig. 9(a). Because, since $\epsilon = 0.1$ is set in this figure, the $\kappa$ axis scale corresponds to the $\kappa\epsilon$ axis scale in Fig. 9(a). This figure shows that $\lambda > 0$ is established in the unstable region of Fig. 9(a).

Fig. 10(d) shows how $\lambda$ in Case 3 depends on $\kappa$ and $\epsilon$, where $N = 0.05$ and $\xi_c = 6$. The line of $\epsilon = 0.1$ is also observed on the back plane of Fig. 10(c), which is drawn at $\xi_c = 6$. As $\epsilon$ increases from 0.05 to 0.2, the $\lambda$ peak becomes higher, and the $\kappa$ value at the maximum $\lambda$ shifts toward $\kappa = 0$, i.e., the left side of this figure. Observing more carefully the line of $\epsilon = 0.1$, the unstable region, i.e., $\lambda > 0$, is located in $0.75 < \kappa < 5.6$. Observing similarly the line of $\epsilon = 0.2$, the unstable region is located in $0.37 < \kappa < 2.8$. Hence, when the unstable region shifts for increasing $\epsilon$ from 0.05 to 0.2, the critical $\kappa\epsilon$ value at $\lambda = 0$ is exactly maintained, as mentioned above. In addition, regarding the shift, when $\epsilon$ increases from 0.1 to 0.2, $\lambda$ at $\kappa = 1$ increases in Fig. 10(d). Meanwhile, $\lambda$ in Case 1 always decreases as $\epsilon$ increases (e.g., Figs. 10(a) and (b) of Shimizu's paper [36]). Hence, the inversion of the $\lambda$ change discussed in Fig. 10(b) is explained by the $\kappa$ shift of the unstable region in Case 3, i.e., viscosity effect.

Fig. 11 shows how the local maximum points of $\Phi$ and $\Psi$ move for $\lambda$ in Cases 1 and 3, where four cases are studied with the combinations for $N = 0$ and 0.05 and for $\epsilon = 0.1$ and 0.2, where $\kappa = 1$ is fixed. The red lines indicate the movement of the local maximum points of $\Phi$, and the black lines indicate that of $\Psi$. The locations of those local maximum points may be convenient for this linear theory to be applied to MHD simulations, rather than $\xi_c$ [36]. As shown in Fig. 11, as those local maximum points are separated from $\xi = 0$, $\lambda$ monotonically increases. In the case of $N = 0$, i.e., Case 1, it seems that those local maximum points do not exceed $\xi = 1.307$, which is the outer edge of the current sheet, i.e., where the increase in $\lambda$ stops. In contrast, in the case of $N = 0.05$, i.e., Case 3, those points can exceed $\xi = 1.307$. In addition, independent of $\epsilon$, as $N$ increases from zero to 0.05, $\lambda$ tends to drastically decrease. It is remarkable that , as $\epsilon$ increases from 0.1 to 0.2, the local maximum point for $N = 0$ slightly shifts to the right but the local maximum point for $N = 0.05$ largely shifts to the left. This inversion of the movements of the local maximum points may be related to Fig. 10(d).

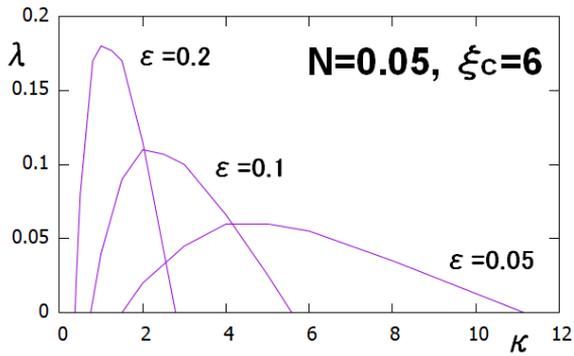

Fig. 10(**d**)Linear growth rate $\lambda$ with respect to $\kappa$ in Case 3 for $\epsilon = 0.05, 0.1, 0.2$, $N = 0.05$, and $\xi_c = 6$.

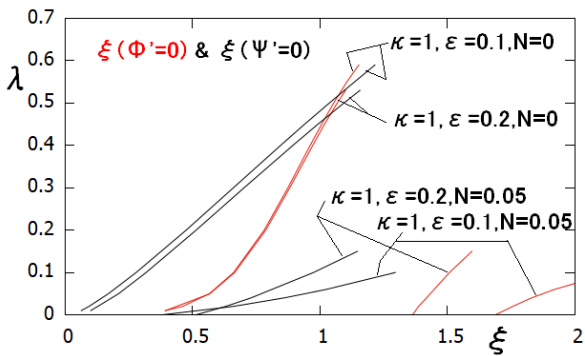

Fig. 11 : Linear growth rate $\lambda$ with respect to the movements of the local maximum points of $\Phi$ and $\Psi$ in Case 3 for $\kappa = 1$ with $\epsilon = 0.1, 0.2$ and $N = 0, 0.05$.

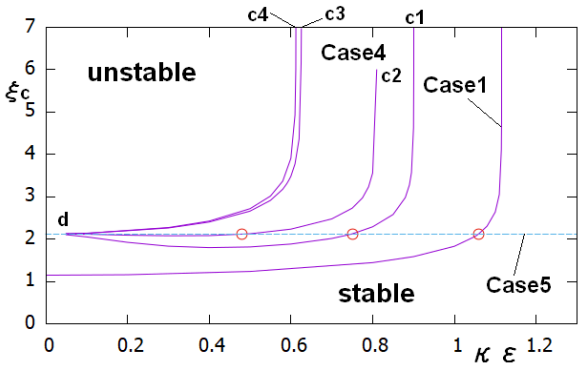

Fig. 12 : Critical conditions of Cases 1, 4, and 5. Case 1 is also plotted in Fig.9(a). Case 4 is for 0.0005(c1-d1), 0.002(c2-d2), 0.05(c3-d3), and $N = +\infty$(c4-d4). Red circles indicate the change point of the solution types defined in Fig. 9(a). Case 5 is independent of $N$.

### 4.3 Cases 4 and 5

Fig. 12 shows the relation of $\kappa\epsilon$ and $\xi_c$ for $\lambda = 0$ of Cases 1a, 4a, and 5a. Hence, the format of Fig. 12 is exactly the same as that of Fig. 9(a), and what are shown in this figure coincide with the foot points of Cases 1 and 4 in Fig. 8. Case 1 in Fig. 12 is the same as that of Fig. 9(a). Above each line is the unstable region of the tearing instability, and below is the stable region. Remarkably, the stable region observed in $\kappa\epsilon < 0.2$ and $\xi_c > 2.12$ of Fig. 9(a) is not observed in Fig. 12. In other words, the parts of d1, d2, d3, and d4 observed in Fig. 9(a) seem to converge to Point d in Fig. 12. This means that the current sheet is not stabilized in $\xi_c > 2.12$ even in $N = +\infty$, because the unstable region steadily exists in $0 < \kappa\epsilon < 0.6$. Another remarkable aspect is that Case 5 observed in Fig. 12 has no stable region in $\xi_c > 2.12$, regardless of $\kappa\epsilon$ and $N$.

Fig. 13(a) shows how $\lambda$ changes for $N$ in various $\kappa$ for Case 4, where some cases of $\xi_c = 3.6$ and 4.8 are studied. Hence, this figure format is almost the same as that of Fig. 10(a). The $\lambda$ value in $N = 0$ is obtained in Case 1. As $N$ increases, $\lambda$ monotonically decreases but does not reach zero. This is evidently different from Fig. 10(a). This means that the zero-crossing solutions exist in $N = +\infty$. In $N > 0.006$, $\lambda$ in every line is almost saturated.

Fig. 13(b) shows how $\lambda$ changes for $\xi_c$ with regards to some $\epsilon$ values for Cases 4b and 4c. Hence, this figure format is the same as that of Figs. 8 and 10(b). As $\epsilon$ approaches zero, $\lambda$ monotonically increases and seems to saturate at a value below unity. The $\lambda$ value in every case studied in this paper does not exceed unity, which is consistent with previous works [36]. This means that the tearing instability always grows at a sub-Alfvenic speed, i.e., $\lambda < 1$.

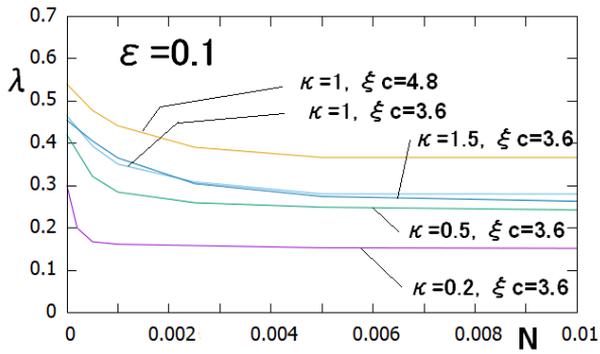

Fig. 13(**a**)Linear growth rate $\lambda$ with respect to $N$ in Case 4 for $\kappa = 0.2, 0.5, 1, 1.5$, $\epsilon = 0.1$, and $\xi_c = 3.6, 4.8$.

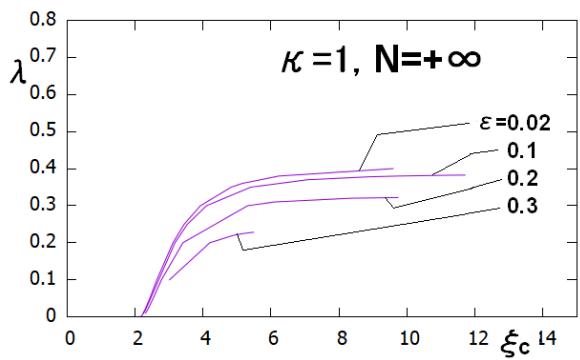

Fig. 13(**b**)Linear growth rate $\lambda$ with respect to $\xi_c$ in Case 4 for $\kappa = 1$, $\epsilon = 0.02, 0.1, 0.2, 0.3$, and $N = +\infty$.

# 5    Discussions

## 5.1    Cases 1 vs 2: Non-uniformity of the Resistivity

First, we compare Cases 1 and 2, in which the linear growth rate $\lambda$ is affected by the non-uniformity of the resistivity. These cases are inviscid. Case 2 is close to the FKR and original LSC theories because the resistivity is ignored in the outer region. In those two theories, the resistivity in the outer region is considered not important to determine the linear growth rate. Hence, it is expected that $\lambda$ in Cases 1 and 2 are the same. As expected, Fig. 8 shows that the $\lambda$'s in Cases 1, 1b and 2 almost coincides. However, Fig. 9(a) shows that the critical condition in $\xi_c > 3$ is separated between Cases 1 and 2. This means that the non-uniformity of the resistivity assists in causing tearing instability. The non-uniformity in Case 2 is just an example. There are a lot of variations for the non-uniformity, such as an anomalous resistivity. This paper suggests that, regarding the speed at which the tearing instability grows, more variations in the non-uniformity of the resistivity must be studied.

## 5.2    Cases 1 vs 3: Introduction of the Viscosity

The viscosity will slowly cause tearing instability. The slowing seems to originate from the change in the upstream open boundary by the introduction of the viscosity. In fact, in Case 1, $\Phi'(\xi_c) = u_y < 0$ is possible at the upstream boundary, resulting in zero-crossing solutions. It means that the plasma flow $u_y$ along the upstream boundary is non-zero, where $u_y < 0$ generates the plasma convection around the boundary, which promotes the plasma inflow toward the X-point. In contrast, in Case 3, $\Phi'(\xi_c) = u_y = 0$ is maintained, resulting in zero-contact solutions. Hence, $\lambda$ in Case 1 is much higher than that of Case 3, as shown in Fig. 8.

Rather, the most remarkable point found in this paper is that, as shown in d1~d4 of Fig. 9(a), the stable region in Case 3 appears in $\kappa\epsilon < 0.2$ and $\kappa\epsilon > 0.45 \sim 0.9$, while the region in Case 1 appears only in $\kappa\epsilon > 1.1$. This difference between Cases 1 and 3 is important. This means that the current sheet in Case 1 is not stabilized in $\xi_c > 1.1$, where tearing instability occurs in any resistivity. Meanwhile, the current sheet in Case 3 is stabilized below a critical $\xi_c$ value that depends on $N$. The critical $\xi_c$ is indicated as the cross points in Fig. 9(a), which are defined as $\xi_{crit}$ and later applied to the MHD simulations.

## 5.3    Case 3 vs 4 and 5: Non-uniformity of the Viscosity

In this section, the non-uniformity of the viscosity is discussed. As shown in Fig. 8, $\lambda$ in Case 3 is much lower than those in Cases 4 and 5. This means that the uniformity of the viscosity steadily suppresses the tearing instability. As mentioned in Section 5.1, regarding the speed at which the tearing instability grows, more variations in the nonuniformity of the viscosity must be studied .

In every finite differential MHD simulation, numerical viscosity and resistivity are implicitly employed to prevent numerical explosions, which are designated to effectively work in an extremely thin current sheet. Such numerical viscosity and resistivity may be non-uniform, which are similar to Cases 2, 4 and 5. In particular, in the MHD simulations of plasmoid instability (PI) for high $S$, as tearing instability is repeated, the current sheet gradually becomes thinner. Then, the tearing instability in the thin current sheet may be artificially faster by the numerical viscosity and resistivity. Since such MHD simulations assume uniform resisitivity, the effect of the numerical non-uniform viscosity and resistivity, i.e., numerical dissipations, must be carefully checked.

## 5.4  Cases 4 and 5: Variations in Non-uniform Viscosities

Regardless of whether the non-uniform viscosity is physical or numerical, when non-uniform viscosity is employed in MHD simulations, either Case 4 or 5 will occur. Otherwise, those cases and Case 3 may mixedly appear. The appearance will depend on the selection of the numerical schemes and the type of boundary conditions in the MHD simulation. In particular, the difference between Cases 4 and 5 is related to whether the differential discontinuity appears at $\xi = 1.307$ or not. In Case 5, the differential discontinuity appears, which may be classified into a mathematically weak solution. In Case 4, it does not appear, which may be classified into a strong solution. Fig. 8 shows that $\lambda$ in Case 4 tends to be higher than those in Cases 3 and 5. That may originate in the difference of the upstream boundary conditions, as mentioned in Section 5.2.

## 5.5  Summary of the Comparisons of Cases 1-5

Eventually, through the comparisons of Cases 1-5, we conclude that the linear growth of the tearing instability is significantly affected by the non-uniformity of the viscosity and the upstream boundary conditions.

## 5.6  Applications

### 5.6.1  Critical Condition $N/\xi_{crit} = 0.06$ of Case 3

This section is the highlight of this paper. Before the applications, we derive the critical condition for Case 3, beyond which the tearing instability occurs and below which the current sheet is stabilized. The cross points in Fig. 9(a) indicate the locations of the critical $\xi_c$ beyond which the unstable $\kappa\epsilon$ range appears. We define the critical $\xi_c$ as $\xi_{crit}$. When $\xi_c > \xi_{crit}$, tearing instability occurs at the largest growth rate in the unstable $\kappa\epsilon$ range. Inversely, when $\xi_c < \xi_{crit}$, since an unstable $\kappa\epsilon$ range does not appear, the current sheet is stable for the tearing mode. As shown in Fig. 9(a), $\xi_{crit}$ depends on $N$. Fig. 14(a) shows how $N$ and $dN/d\xi_{crit}$ depend on $\xi_{crit}$, where $N$ monotonically increases with respect to $\xi_{crit}$. For larger $\xi_{crit}$, $dN/d\xi_{crit}$ seems to be almost saturated at 0.06. We assume that the increase in $N$ is linear. In that case, as shown by the dashed line in Fig. 14(a), the linear increase is approximately measured as $N = 0.06(\xi_{crit} - 2.65)$, which may be simplified to $N = 0.06\xi_{crit}$ for sufficiently large $\xi_{crit}$. As a result, tearing instability occurs in $N/\xi_{crit} = 0.06 > N/\xi_c$, and the current sheet is stable in $N/\xi_c > 0.06$. It is remarkable that since $\xi_{crit}$ in Cases 1 and 2 is fixed at 1.1 and those in Cases 4 and 5 are fixed at 2.1, the current sheet is not stabilized at larger $\xi_c$. This means that the uniform viscosity assumed in Case 3 is important to stabilize the current sheet. Next, we discuss the meaning of $N/\xi_{crit} = 0.06$ in the MHD simulations of PI.

### 5.6.2  Measurement of $\xi_c$ and $N$ in MHD simulations

Simply, $\xi_c$ is the location of the upstream open boundary which will be close to the open boundary often employed in MHD simulations. We define the location in the MHD simulations as $L_x$, which is the size of the simulation box in the normal direction to the current sheet [36,54]. Then, it results in $\xi_c = 1.307L_x/\delta_{cs}$, where the outer edge of the current sheet in the modified LSC theory is fixed at $\xi_0 = 1.307$ and the outer edge in the MHD simulations is $\delta_{cs}$. In most MHD simulations, $L_x$ is fixed throughout the execution of the simulation. However, $\delta_{cs}$ may change, depending on each single event of the tearing instability repeated in PI. Hence, $\xi_c = 1.307L_x/\delta_{cs}$ may change in PI and is interpreted to be the inversion of $\delta_{cs}$, which is related to the resistivity. On the other hand, in most MHD simulations of PI, the resistivity $\eta$ and viscosity $\nu$ are maintained as constant throughout the execution of the simulation. These values are translated to $\epsilon$ and $N$, respectively, via $\epsilon^2/2 = \eta/(V_A L_{cs})$ and $N = 2\nu/(V_A L_{cs})$, where $\epsilon = 2\delta_{cs}/L_{cs}$. In most MHD simulations of PI, $V_A$ almost does not change throughout the execution, but $L_{cs}$ may change in each single event of the tearing instability [36]. Hence, the viscosity $N = 2\nu/(V_A L_{cs})$ is interpreted to be the inversion of $L_{cs}$, which is related to the viscosity. As a result, Fig. 14(a) shows how $\delta_{cs}$ and $L_{cs}$ change in the critical condition.

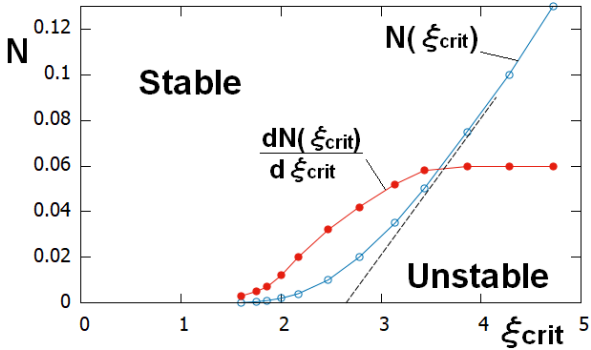

Fig. 14(**a**): Relation of $N$ and $\xi_{crit}$ in Case 3, which is obtained from Fig. 9(a). This figure shows when the current sheet is stabilized, i.e., the critical condition.

Eventually, the linear critical condition $N/\xi_c = 0.06$ is translated to $\eta\nu/(1.307V_A^2 L_x \delta_{cs}) = 0.06$ in the MHD simulation. In addition, by defining the Lundquist number $S = V_A L_{cs}/\eta$ and magnetic Prandtl number $P_m = \nu/\eta$, $N/\xi_c = 2P_m/(S\xi_c)$ is obtained. Since $1.307/\xi_c$ is the ratio of the current sheet thickness and the distance between the upstream open boundary and the neutral sheet, $2P_m/(S\xi_c)$ consists of those three dimensionless parameters.

### 5.6.3 Forward and Inverse Cascade Processes of PI

There are many MHD numerical studies on the cascade process observed in PI, where the tearing instability is intermittently repeated and the size is gradually changed from a large scale to a small scale [20,22,25,26,27,28,29,30,42,43, 44]. The scale change is caused by the forward cascade process. If the forward cascade is sufficiently developed, the current sheet will become a turbulent state. Those studies expect that a few large-scale (monster) plasmoids are formed at a fairly high magnetic reconnection rate [31,32,33,45,46]. To generate such large-scale plasmoids, the forward cascade process must be stopped or sufficiently suppressed. Because, the forward cascade cannot generate such large-scale plasmoids. Once the forward cascade occurs, the formation of large-scale plasmoids will require the inverse cascade process by which the magnetic energy is converted from a small scale to a large scale [47,48]. However, it is still unclear how the forward cascade process observed in PI is stopped [7,29,49] or suppressed in the extremely thin current sheet.

### 5.6.4   Limit of the Forward Cascade

Since the cascade process is essentially non-linear, the application of the modified LSC theory, which is linear, must be limited. In this section, we focus only on how the forward cascade stops in PI, for which linear theory will be applicable. When the forward cascade sufficiently proceeds, the tearing instability will eventually occur at an extremely small scale. Many people will expect that such an extremely small-scale tearing instability somehow stops, i.e., is stabilized. It should certainly stop, but it is unclear how the forward cascade stops in the MHD simulations of PI. The issue must be considered.

First, the tearing instability on a smaller scale than the numerical grid size cannot be correctly studied in finite differential MHD simulations. In that case, numerical explosions will occur, because of numerical errors. To prevent the numerical explosion, numerical dissipations may be effectively implemented, in which any artificial resistivity and viscosity should be included. In such cases, the numerical result will depend on the selection of the numerical scheme, such as HLLD or 2-step Lax-Wendroff, where the resistivity and viscosity will no longer be uniform. If so, PI activated by "uniform" resistivity cannot be exactly studied in such simulations. To focus on the uniform resistivity and viscosity, the forward cascade must stop with the uniform dissipations before the scale reaches the numerical grid size.

Second, if the forward cascade does not stop, an inverse cascade may not occur. In that case, very large plasmoids will not be generated, which is expected for PI to be a fast magnetic reconnection process. Hence, it should be worth studying how the forward cascade stops in PI.

### 5.6.5   Critical Condition in the MHD Simulations of PI

In this section, we select two previous MHD simulations [36,54] and apply the critical condition $\eta\nu/(1.307V_A^2 L_x \delta_{cs}) = 0.06$ to them.

The MHD simulation [54] executed by HLLD scheme is for inviscid-resistive MHD, i.e., the case of $\nu = 0$. Similarly, there are many numerical MHD studies of PI in which the viscosity [7,36,50,53] is incorporated. In those cases, since $\eta\nu/(1.307V_A^2 L_x \delta_{cs}) = 0.0$, the forward cascade does not stop, as shown in Fig. 9(a). In fact, the numerical dissipations implicitly included in the numerical scheme will effectively work to stop the forward cascade and prevent numerical explosion. Since such numerical dissipation will have a non-uniform resistivity, fully developed PI activated by uniform resistivity cannot be correctly explored. Fortunately, the MHD simulation executed by Shimizu [54] was only aimed at exploring how the forward cascade starts. Hence, how the forward cascade stops was not explored.

The MHD simulation [36] executed by 2-step Lax-Wendroff scheme is for viscous-resistive MHD, where $\eta = 0.016$, $\nu = 0.008$, $V_A = 5.5$, and $L_x = 200$. Rigorously, since the upstream open boundary at $x = L_x$ has a finite magnetic field intensity, the boundary is different from that of the modified LSC theory. In fact, the magnetic field intensity $f(\xi)$ is close to zero at large $\xi_c$, as shown in Figs. 1-5. The difference may be problematic. However, let us show an application as an example. We directly obtain $\eta\nu/(1.307V_A^2 L_x \delta_{cs}) = 1.6 \times 10^{-8}/\delta_{cs}$. In the simulation, the thickness $\delta_{cs}$ of the current sheet is initially 1.0. As PI proceeds, $\delta_{cs}$ gradually decreases from 1.0. For example, $\delta_{cs} = 0.5$ is observed in the secondary tearing instability. Finally, we stopped the simulation when the fourth tearing instability started. Hence, the simulation was not executed until the forward cascade stops, where the inverse cascade and monster plasmoids were not observed. If the simulation were continued, the smallest $\delta_{cs}$ may reach the numerical grid size $\Delta x = 0.02$, where the current sheet can no longer be treated in the simulation, i.e., the simulation fails. In the extreme case of $\delta_{cs} = \Delta x = 0.02$, since $\eta\nu/(1.307V_A^2 L_x\delta_{cs}) = 8.1 \times 10^{-7} << 0.06$, the forward cascade does not yet stop. This means that the resistivity $\eta$ or viscosity $\nu$ is too small to stop it. By downsizing the simulation box, $L_x$ may be largely reduced from 200 to 2. Even in that case, it does not stop. To steadily stop it, $\eta\nu$ must be roughly increased by $10^5$ times. Otherwise, $\Delta x$ must be largely reduced., i.e., a much higher numerical resolution is required to observe when the forward cascade stops.

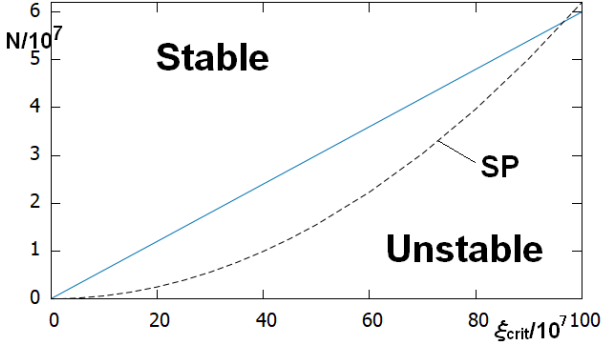

Fig. 14(**b**): Relation of $N$ and $\xi_{crit}$ in Case 3. The blue solid line is extrapolated from Fig. 14(a), where the axis scales are changed by $10^7$ times. Hence, Fig. 14(a) is localized around the origin of this figure. Dash line SP shows the relation of $\xi_c$ and $N$ in the SP model.

Fig. 14(b) shows the physical meaning of what was discussed here. The range of Fig. 14(a) is extremely localized at the origin of Fig. 14(b). In fact, the oblique straight line shows the critical condition $N = 0.06(\xi_{crit} - 2.65) \sim 0.06\xi_{crit}$ predicted from Fig. 14(a). Then, the quadratic curve is based on the SP model, i.e., $\eta L_{cs} = 2V_A\delta_{cs}^2$. This means that when $V_A$ and $\eta$ are constant, $L_{cs}$ is proportional to $\delta_{cs}^2$. Note that Fig. 14 shows the critical relation of $1/L_{cs}$ and $1/\delta_{cs}$. Hence, the quadratic curve shows that when $1/\delta_{cs}$ slowly becomes larger in the progress of the forward cascade, $1/L_{cs}$ rapidly becomes larger. Hence, the straight line of $N = 0.06(\xi_{crit} - 2.65)$ will cross the curve at a $\xi_{crit}$ value, i.e., around the upper-right corner of Fig. 14(b), where the forward cascade stops.

The discussions in this section may have some problems to be solved in the future. First, the modified LSC theory is based on incompressible MHD while those MHD simulations are on compressible MHD. Hence, this discussion in this section may be affected by the compressibility. Second, as will be mentioned in Section 5.8, the modified LSC theory in the small $\kappa$ range may be modified by the improvement of the WKB approximation. Third, in the extension from Fig. 14(a) to (b), the linearity of $N - \xi_{crit}$ was assumed.

## 5.7   Application for Real Plasma Physics

It will be important to discuss the trigger problem of the tearing instability observed in real plasma observations, such as substorms and solar flares. In general, we may expect that, as the current sheet becomes thin, the tearing instability is initiated at a certain time. As discussed in the preceding section, the thinning increases $\xi_c$. If $N$ is assumed to be constant in the thinning, the operating point $(N, \xi_c)$ located in Figs. 14(a) and (b) moves horizontally from left to right. If the operating point moves from the stable region to the unstable region, the tearing instability is initiated there.

Unfortunately, it is difficult to measure $N$ and $\xi_c$ in a real current sheet. The largest problem is that the equilibrium in the modified LSC theory is 2D but that in real plasma phenomena will be established in any 3D structure, e.g., the magnetosphere and magnetic flux tube. Since the 3D structure in real plasma will always have a finite size, a finite $\xi_c$ value will actually exist at an upstream point in the 3D structure. Hence, Figs. 14(a) and (b) show the applicability of the trigger problem in real plasma observations.

## 5.8   Comparison with previous linear theories

There are many studies on the linear theory of tearing instabilities, where the viscosity was studied [5,6,9,11,12,13,21, 22,24, 51,55]. In most of these studies, the traditional FKR analysis style, in which the outer region is assumed to be an ideal MHD and is connected to the resistive MHD inner region with the $\Delta'$ index, is implemented. Hence, they may be close to Cases 4 and 5. In addition, except for the employment of a periodic boundary, the upstream boundary is set at the infinity point $\xi_c = +\infty$. In this paper, we suggest that the forward cascade does not stop in those cases. In other words, tearing instability occurs even in the unlimitedly thin current sheet. This may be consistent with Porcelli's study [6], where it was reported that the viscosity slows the tearing instability but does not remove it. In other words, it may have a delicate problem of the double limit of $\xi_c = +\infty$ and $N = +\infty$. At least, Case 3 in this paper showed that the viscosity in the outer region is required to stop it. On the other hand, Loureiro's study [21] may be similar to Case 3. In fact, in their study, the open boundary condition of $\Phi''(\xi_c) = 0$ was employed at a finite $\xi_c < +\infty$. However, it seems that the critical condition of $\lambda = 0$, such as those shown in Figs. 9(a), and hence, 14(a), was not obtained in their paper.

## 5.9   WKB Approximation

The modified LSC theory developed in this paper is based on the linear perturbation equations studied in the original LSC theory [8]. When the equations were introduced in the original LSC theory, the lowest order of the WKB approximation was taken. For that reason, the modified LSC theory shown in this paper is inapplicable for the low $\kappa$ range, i.e., $\kappa \sim 0$. To explore the tearing instability in the low $\kappa$ range, a higher-order WKB approximation must be made. The improvement of the WKB approximation will be the next work to be studied (Eqs. (23) and (24) in Appendix, and [41]).

# 6   Summary

In this paper, the linear theory of the tearing instability in inviscid-resistive MHD was extended to viscous-resistive MHD. The largest difference from previous studies of linear theory is the introduction of the upstream open boundary at a finite point. As listed in Table 1. in addition to when the resistivity and viscosity are uniform, some variations in the non-uniformity were also studied. When the non-uniformity is introduced in the resistivity and viscosity, i.e., Cases 2, 4, and 5, it was shown that the linear growth rate $\lambda$ can be enhanced, rather than the uniform case. This also means that the tearing instability can be caused in an unlimitedly thin current sheet. It is remarkable that, even in the infinite viscosity $N = +\infty$, Cases 4 and 5 are not stabilized. Hence, in those non-uniform cases, the forward cascade of plasmoid instability (PI) does not stop. This suggests that the MHD equation for those non-uniform cases fails to fully describe the end of the forward cascade. Such an MHD equa-tion cannot be exactly solved in finite differential numerical simulations because instability occurs on a smaller scale than the numerical grid sizes, i.e., $\Delta x$.

In addition, in Case 3, it was also shown that, in $N/\xi_c = \eta\nu/(1.307 V_A^2 L_x \delta_{cs}) > 0.06$, the current sheet is stabilized for the tearing mode, i.e., the forward cascade stops. Then, the critical condition was applied to two MHD simulations [36,54]. It suggests that the viscosity $\nu$ and resistivity $\eta$ assumed in the high-S MHD simulations of PI is too small to stop the forward cascade before the thickness of the current sheet reaches the numerical grid size.

# 7   Acknowledgments

This study was supported by MEXT/JSPS KAKENHI under Grant No. 21K03645. The numerical calculations were performed on parallel computer systems at Kyoto and Nagoya University Data Processing Centers. The data analysis executed in this study was partially assisted by Mr. Homma and Mr. Nishimura who were undergraduate students in the Research Center for Space and Cosmic Evolution (RCSCE) of Ehime University. The author thanks them for their assistance.

# 8   Appendix

In this section, following Loureiro's notation [8], we introduce viscosity to the inviscid perturbation equations proposed by Loureiro. First, we employ the equilibrium of $\phi_0$ and $\psi_0$ derived in Chapter 2. Then, we set $\phi = \phi_0 + \delta\phi$, $\psi = \psi_0 + \delta\psi$, where $\delta\phi(x, y, t) = \phi_1(x, t)e^{ik(t)y}$ and $\delta\psi(x, y, t) = \psi_1(x, t)e^{ik(t)y}$ with $k(t) = k_0 e^{-\Gamma_0 t}$. The perturbation equations for Eqs. (3) and (4) are translated as follows:

$$(\partial_x^2 - k^2)\partial_t\phi_1 - \Gamma_0 x\partial_x(\partial_x^2 - k^2)\phi_1 + 2\Gamma_0 k^2\phi_1$$
$$= [B_{0y}(x)(\partial_x^2 - k^2) - \partial_x^2 B_{0y}(x)]ik\psi_1 + \nu(\partial_x^4 - 2k^2\partial_x^2 + k^4)\phi_1 \tag{21}$$

$$\partial_t\psi_1 - \Gamma_0 x\partial_x\psi_1 - B_{0y}ik\phi_1 = \eta(\partial_x^2 - k^2)\psi_1 \tag{22}$$

Except for the viscosity term, these equations are exactly the same as Eqs. (6) and (7) that were derived from the original LSC theory [8]. Furthermore, we set $\phi_1 = -i\Phi(x)e^{\gamma t}$ and $\psi_1 = \Psi(x)e^{\gamma t}$. In the same manner as the derivation of the equilibrium, we translate Eqs. (21) and (22) with $x = \delta_{cs}\xi$, $\kappa = k_0 V_A/\Gamma_0 = k_0 L_{cs}/2$, $\epsilon = 2\delta_{cs}/L_{cs}$, $\lambda = \gamma/(\Gamma_0\kappa)$ and $N = 2\nu/(V_A L_{cs}) = \nu/\Gamma_0$.

$$N\Phi'''' = \kappa^2\epsilon^2((\lambda + 2\kappa N)\Phi'' - (\lambda + \kappa N)\kappa^2\epsilon^2\Phi + f(\xi)(\Psi'' - \kappa^2\epsilon^2\Psi) - f''(\xi)\Psi$$
$$-\xi\Phi'''/\kappa + \kappa\epsilon^2\xi\Phi' + 2\kappa\epsilon^2\Phi) \tag{23}$$

$$\Psi'' - \kappa^2\epsilon^2\Psi = \kappa\lambda\Psi - \kappa f(\xi)\Phi - \xi\Psi' \tag{24}$$

where $f(\xi) = B_{0y}/V_A$. Regarding the tearing instability, $\Phi > 0$ and $\Psi > 0$ must be required for $\xi > 0$. The prime of $\Phi$ and $\Psi$ indicates the derivative with respect to $\xi$, which is normalized by the characteristic width $\delta_{cs}$ of the current sheet, as $\xi = x/\delta_{cs}$. Following the LSC notation, $\lambda$ is the growth rate normalized by $l_{cs}/V_A$, where $l_{cs}$ is the wavelength of the plasmoid chain and $V_A$ is the Alfven speed measured in the upstream magnetic field region. $\kappa = \pi L_{cs}/l_{cs}$ is the wavenumber along the current sheet, where $L_{cs}$ is the total length of the steady-state SP sheet. As mentioned in Shimizu's paper [36], $L_{cs}$ is related to the spatial gradient of the outflow velocity $u_{y0}$ measured at the X-point, as $du_{y0}/dy = 2V_A/L_{cs}$.

In fact, Eqs.(23) and (24) can be numerically solved by the IVP technique but it is fairly difficult. As mentioned in Section 5.8, this will be examined in future works. For simplicity, in this paper, $\gamma >> \Gamma_0$, i.e., $k(t) = k_0$, is assumed. This simplification is the same as that in the original LSC theory [8]. Then, Eqs. (23) and (24) are transferred as follows.

$$N\Phi'''' = \kappa\epsilon^2((\lambda + 2\kappa N)\Phi'' - (\lambda + \kappa N)\kappa^2\epsilon^2\Phi + f(\xi)(\Psi'' - \kappa^2\epsilon^2\Psi) - f''(\xi)\Psi) \qquad (25)$$

$$\Psi'' - \kappa^2\epsilon^2\Psi = \kappa\lambda\Psi - \kappa f(\xi)\Phi \qquad (26)$$

These equations are solved in Cases 3, 4, and 5. These equations with N=0 are solved in Cases 1 and 2. Here, the assumption of $\gamma >> \Gamma_0$ means that $k(t)$ slowly changes in the linear growth time. Notably, as $y$ increases, since the plasma outflow speed $u_{0y} = \partial_x\phi_0 = \Gamma_0 y$ unlimitedly increases, $k(t)$ quickly changes over time. Hence, this assumption is applicable in $y \sim 0$ and $k >> 0$.

[1]  E.N.Parker, J. Geophys. Res. **62**, 509 (1957).

[2]  H.P.Fruth, J.Killeen, and M.N.Rosenbruth, Phys. Fluids **Vol.6** No.4, 459 (1963).

[3]  Petschek,H.E., AAS-NASA Symposium on the Physics of Solar Flares, NASA Special Publ. **SP-50**, 425 (1964).

[4]  V.M.Vasyliunas, Rev. Geophys. Space Phys. **13**, 303 (1975).

[5]  A. Bondeson and M. Persson, Phys. Fluids **29(9)**, 2997 (1986).

[6]  F. Porcelli, Phys. Fluids **30(6)**, 1734 (1987).

[7]  A.Tenerani, M.Velli, A.F.Rappazzo, and F.Pucci, Astro.Phys.J. Lett., **813**:L32 (2015).

[8]  N.F.Loureiro, A.A.Schekochihin, and S.C.Cowley, Phys. Plasmas **14** 100703 (2007).

[9]  G. Einaudi and F. Rubini, Phys. Fluids **B1(11)**, 2224 (1989).

[10]  Pucci and Velli, Astro.Phys.J.Lett., 780 L19 (2014),

[11]  A.D.Wood, E.O'riordan, N.Sweeney, and R.B.Paris, J.Plasma Phys., **Vol.70, part 2** 155 (2004).

[12]  L.Ofman, X.L.Chen, P.J.Morrison, and R.S.Steinolfson, Phys. Fluids **B3(6)** 1364 (1991).

[13]  D. Grasso, R.J.Hastie, F.Porcelli, and C.Tebaldi, Phys. Plasmas **15** 072113 (2008).

[14]  M.Ugai and T.Tsuda, J. Plasma Phys. **17**, 337 (1977).

[15]  M.Ugai, Plasma Phys. and Controlled Fusion **26**, 12B, 1549 (1984).

[16]  R.M.Kulsrud, Earth Planets Space **53**, (2001).

[17]  T.Shimizu, and M.Ugai, Phys. Plasmas **4**, 921 (2003).

[18]  H.Baty, E.R.Priest, and T.G.Forbes, Phys. Plasmas **13**, 022312 (2006).

[19]  R.M. Kulsrud, Phys. Plasmas **18**, 111201 (2011).

[20]  R.Samtaney, N.F.Loureiro, D.A.Uzdensky, A.A.Schekochihin, and S.C.Cowley, Phys.Rev.Lett. **103**, 105004 (2009).

[21]  N.F.Loureiro, A.A.Schekochihin, and D.A.Uzdensky, Phys. Rev. **E87** 012102 (2013).

[22]  L.Comisso, and D.Grasso, Phys. Plasmas **23** 032111 (2016).

[23]  Biskamp,D., Phys. Fluids **29**, 1520 (1986).

[24]  R.D. Parker, R.L.Dewar, and J.L.Johnson, Phys. Fluids **B2(3)** 508 (1990).

[25]  N.F.Loureiro, D.A.Uzdensky, A.A.Schekochihin, S.C.Cowley, and T.A.Yousef, Mon.Not.R.Astron.Soc. **399**, L146-L150 (2009).

[26]  A.Bhattacharjee, Y.M.Huang, H.Yang, and B.Rogers, Phys. Plasmas **16**, 112102 (2009).

[27]  P.A.Cassak, and J.F.Drake, Astro.Phys.J. **707**, L158-L162 (2009).

[28]  L.Ni, U.Ziegler, Yi-Min.Huang, J.Lin, and Z.Mei, Phys. Plasmas **19**, 072902 (2012).

[29]  Y.M.Huang, and A.Bhattacharjee, Phys.Plasmas **20**, 055702 (2013).

[30]  H.Baty, J.Plasma Phys. **vol.80, part 5**, 655-665 (2014).

[31]  M.J.Nemati, Z.X.Wang, and L.Wei, Astro.Phys.J., 821,128 (2016)

[32]  N.F.Loureiro, R.Samtaney, A.A.Schekochihin, and D.A.Uzdensky, Phy.Plasmas, Vol.19,4,042303 (2012).

[33]  D.A.Uzdensky,N.F.Loureiro, and A.A.Schekochihin, Phys.Rev.Lett., 105, 235002 (2010).

[34]  T.Shimizu, KDK Research Report 2017, RISH Kyoto University Japan, 87-90 (2018).

[35]  T.Shimizu, 2nd Asia-Pacific Conf. on Plasma Phys., Kanazawa, Japan, SGP-04 (2018).

[36]  T.Shimizu, http://arxiv.org/abs/2209.00149 (2022).

[37]  T.Shimizu, KDK Research Report 2019, RISH Kyoto University Japan, 73-76 (2020).

[38]  T.Shimizu, 5th Asia-Pacific Conf. on Plasma Phys., Fukuoka, Japan, SG-I41 (2021).

[39]  T.Shimizu, KDK Research Report 2021, RISH Kyoto University, Japan, 93-96 (2022).

[40]  T.Shimizu, 6th Asia-Pacific Conf. on Plasma Phys., Nagoya, Japan, SGP-10 (2022).

[41]  T.Shimizu, KDK Research Report 2022, RISH Kyoto University Japan, 53-60 (2023).

[42]  L.Comisso, Y.-M.Huang, M.Lingam, E.Hirvijoki,and, A.Bhattacharjee, arXiv:1802.02256v2 (2018).

[43]  C.Dong, L.Wang, Y.-M.Huang, L.Comisso, and A.Bhattacharjee, Phys.Rev.Lett., 121, 165101 (2018).

[44]  C.Dong, L.Wang, Y.-M. Huang, L.Comisso, and A. Bhattacharjee, Science Advances, Vol 8, Issue 49, DOI:10.1126/sciadv.abn7627 (2022)

[45]  H.Lotfi and M.Hosseinpour, Frontier in Astronomy and Space Science, Vol.8, Article 768965, (2021)

[46]  N.F.Loureiro and D.A.Uzdensky, Plasma Phys. Control Fusion, 58, 014021 (2016)

[47]  W.C.Muller, S.K.Malapaka, and A.Busse, Phys.Rev., E85, 015302 (2012)

[48]  N.T. Baker, A. Potherat, L. Davoust, and F.Debray, https://arxiv.org/pdf/1708.03494.pdf (2018)

[49]  Y.M.Huang, L.Comisso, and A.Bhattacharjee, Astro Phys J., 849:75(18pp), (2017).

[50]  L.N.Wu and Z.W.Ma, Phys. Plasmas **21** 072105 (2014).

[51]  A.Tenerani, A.F.Rappazzo, M.Velli, and F.Pucci, Astro.Phys.J., textbf801:145 (2015).

[52]  L.Comisso, and A.Bhattacharjee, J.Plasma Phys., **Vol.82**, Journal Issue 06 (2016).

[53]  T.Minoshima, T.Miyoshi, and S.Imada, Phys. Plasmas **23** 072122 (2016).

[54]  T.Shimizu, K.Kondoh, and S.Zenitani, Phys. Plasmas **24**, 112117 (2017).

[55]  Y.M.Huang, and A.Bhattacharjee, Astro.Phys.J., **818**:20(11pp) (2016).